\newcount\mgnf\newcount\tipi\newcount\tipoformule
\newcount\aux\newcount\piepagina\newcount\xdata
\mgnf=0
\aux=1           
\tipoformule=1   
\piepagina=1     
\xdata=0         
\def\Di{1 settembre 1997}

\ifnum\mgnf=1 \aux=0 \tipoformule =1 \piepagina=1 \xdata=1\fi
\newcount\bibl
\ifnum\mgnf=0\bibl=0\else\bibl=1\fi
\bibl=0

%
%
%
%
\openout8=ref.b
\ifnum\bibl=0
\def\ref#1#2#3{[#1#2]\write8{#1@#2}}
\def\rif#1#2#3#4{\item{[#1#2]} #3}
\fi

\ifnum\bibl=1
\def\ref#1#2#3{[#3]\write8{#1@#2}}
\def\rif#1#2#3#4{}

\fi

\def\9#1{\ifnum\aux=1#1\else\relax\fi}
\ifnum\piepagina=0 \footline={\rlap{\hbox{\copy200}\
$\st[\number\pageno]$}\hss\tenrm \foglio\hss}\fi \ifnum\piepagina=1
\footline={\rlap{\hbox{\copy200}} \hss\tenrm \folio\hss}\fi
\ifnum\piepagina=2\footline{\hss\tenrm\folio\hss}\fi

\ifnum\mgnf=0 \magnification=\magstep0
\hsize=13.5truecm\vsize=22.5truecm \parindent=4.pt\fi
\ifnum\mgnf=1 \magnification=\magstep1
\hsize=16.0truecm\vsize=22.5truecm\baselineskip14pt\vglue5.0truecm
\overfullrule=0pt \parindent=4.pt\fi

\let\a=\alpha\let\b=\beta \let\g=\gamma \let\d=\delta
\let\e=\varepsilon  \let\h=\eta
\let\th=\vartheta \let\l=\lambda \let\m=\mu \let\n=\nu
 \let\p=\pi \let\r=\rho \let\s=\sigma \let\t=\tau
 \let\f=\varphi \let\ps=\psi \let\o=\omega
 \let\G=\Gamma \let\D=\Delta \let\Th=\Theta
\let\L=\Lambda\let\X=\Xi   
  
{\count255=\time\divide\count255 by 60 \xdef\oramin{\number\count255}
\multiply\count255 by-60\advance\count255 by\time
\xdef\oramin{\oramin:\ifnum\count255<10 0\fi\the\count255}}
\def\ora{\oramin }

\ifnum\xdata=0
\def\data{\number\day/\ifcase\month\or gennaio \or
febbraio \or marzo \or aprile \or maggio \or giugno \or luglio \or
agosto \or settembre \or ottobre \or novembre \or dicembre
\fi/\number\year;\ \ora}
\else
\def\data{\Di}
\fi

\setbox200\hbox{$\scriptscriptstyle \data $}
\newcount\pgn \pgn=1
\def\foglio{\number\numsec:\number\pgn
\global\advance\pgn by 1} \def\foglioa{A\number\numsec:\number\pgn
\global\advance\pgn by 1}
\global\newcount\numsec\global\newcount\numfor \global\newcount\numfig
\gdef\profonditastruttura{\dp\strutbox}
\def\senondefinito#1{\expandafter\ifx\csname#1\endcsname\relax}
\def\SIA #1,#2,#3 {\senondefinito{#1#2} \expandafter\xdef\csname
#1#2\endcsname{#3} \else \write16{???? ma #1,#2 e' gia' stato definito
!!!!} \fi} \def\etichetta(#1){(\veroparagrafo.\veraformula) \SIA
e,#1,(\veroparagrafo.\veraformula) \global\advance\numfor by 1
\9{\write15{\string\FU (#1){\equ(#1)}}} \9{ \write16{ EQ \equ(#1) == #1
}}} \def \FU(#1)#2{\SIA fu,#1,#2 }
\def\etichettaa(#1){(A\veroparagrafo.\veraformula) \SIA
e,#1,(A\veroparagrafo.\veraformula) \global\advance\numfor by 1
\9{\write15{\string\FU (#1){\equ(#1)}}} \9{ \write16{ EQ \equ(#1) == #1
}}} \def\getichetta(#1){Fig.  \verafigura \SIA e,#1,{\verafigura}
\global\advance\numfig by 1 \9{\write15{\string\FU (#1){\equ(#1)}}} \9{
\write16{ Fig.  \equ(#1) ha simbolo #1 }}} \newdimen\gwidth \def\BOZZA{
\def\alato(##1){ {\vtop to \profonditastruttura{\baselineskip
\profonditastruttura\vss
\rlap{\kern-\hsize\kern-1.2truecm{$\scriptstyle##1$}}}}}
\def\galato(##1){ \gwidth=\hsize \divide\gwidth by 2 {\vtop to
\profonditastruttura{\baselineskip \profonditastruttura\vss
\rlap{\kern-\gwidth\kern-1.2truecm{$\scriptstyle##1$}}}}} }
\def\alato(#1){} \def\galato(#1){}
\def\veroparagrafo{\number\numsec}\def\veraformula{\number\numfor}
\def\verafigura{\number\numfig}
\def\geq(#1){\getichetta(#1)\galato(#1)}
\def\Eq(#1){\eqno{\etichetta(#1)\alato(#1)}}
\def\eq(#1){\etichetta(#1)\alato(#1)}
\def\Eqa(#1){\eqno{\etichettaa(#1)\alato(#1)}}
\def\eqa(#1){\etichettaa(#1)\alato(#1)}
\def\eqv(#1){\senondefinito{fu#1}$\clubsuit$#1\write16{No translation
for #1} \else\csname fu#1\endcsname\fi}
\def\equ(#1){\senondefinito{e#1}\eqv(#1)\else\csname e#1\endcsname\fi}
\openin13=#1.aux \ifeof13 \relax \else \input #1.aux \closein13\fi
\openin14=\jobname.aux \ifeof14 \relax \else \input \jobname.aux
\closein14 \fi \9{\openout15=\jobname.aux} \newskip\ttglue


\font\titolo=cmbx10 scaled \magstep1
\font\ottorm=cmr8\font\ottoi=cmmi7\font\ottosy=cmsy7
\font\ottobf=cmbx7\font\ottott=cmtt8\font\ottosl=cmsl8\font\ottoit=cmti7
\font\sixrm=cmr6\font\sixbf=cmbx7\font\sixi=cmmi7\font\sixsy=cmsy7

\font\fiverm=cmr5\font\fivesy=cmsy5\font\fivei=cmmi5\font\fivebf=cmbx5
\def\ottopunti{\def\rm{\fam0\ottorm}\textfont0=\ottorm%
\scriptfont0=\sixrm\scriptscriptfont0=\fiverm\textfont1=\ottoi%
\scriptfont1=\sixi\scriptscriptfont1=\fivei\textfont2=\ottosy%
\scriptfont2=\sixsy\scriptscriptfont2=\fivesy\textfont3=\tenex%
\scriptfont3=\tenex\scriptscriptfont3=\tenex\textfont\itfam=\ottoit%
\def\it{\fam\itfam\ottoit}\textfont\slfam=\ottosl%
\def\sl{\fam\slfam\ottosl}\textfont\ttfam=\ottott%
\def\tt{\fam\ttfam\ottott}\textfont\bffam=\ottobf%
\scriptfont\bffam=\sixbf\scriptscriptfont\bffam=\fivebf%
\def\bf{\fam\bffam\ottobf}\tt\ttglue=.5em plus.25em minus.15em%
\setbox\strutbox=\hbox{\vrule height7pt depth2pt width0pt}%
\normalbaselineskip=9pt\let\sc=\sixrm\normalbaselines\rm}
\catcode`@=11
\def\footnote#1{\edef\@sf{\spacefactor\the\spacefactor}#1\@sf
\insert\footins\bgroup\ottopunti\interlinepenalty100\let\par=\endgraf
\leftskip=0pt \rightskip=0pt \splittopskip=10pt plus 1pt minus 1pt
\floatingpenalty=20000
\smallskip\item{#1}\bgroup\strut\aftergroup\@foot\let\next}
\skip\footins=12pt plus 2pt minus 4pt\dimen\footins=30pc\catcode`@=12
\let\nota=\ottopunti\newdimen\xshift \newdimen\xwidth \newdimen\yshift
\def\ins#1#2#3{\vbox to0pt{\kern-#2 \hbox{\kern#1
#3}\vss}\nointerlineskip} \def\eqfig#1#2#3#4#5{ \par\xwidth=#1
\xshift=\hsize \advance\xshift by-\xwidth \divide\xshift by 2
\yshift=#2 \divide\yshift by 2 \line{\hglue\xshift \vbox to #2{\vfil #3
\includegraphics{#4.ps} }\hfill\raise\yshift\hbox{#5}}} \def\8{\write13}

\def\didascalia#1{\vbox{\nota\0#1\hfill}\vskip0.3truecm}
\def\V#1{{\,\underline#1\,}}
\def\T#1{#1\kern-4pt\lower9pt\hbox{$\widetilde{}$}\kern4pt{}}
\let\dpr=\partial \let\io=\infty\let\ig=\int
\def\fra#1#2{{#1\over#2}}\let\0=\noindent
\def\guida{\leaders\hbox to 1em{\hss.\hss}\hfill}
\def\tende#1{\vtop{\ialign{##\crcr\rightarrowfill\crcr
\noalign{\kern-1pt\nointerlineskip} \hglue3.pt${\scriptstyle
#1}$\hglue3.pt\crcr}}} \def\otto{\
{\kern-1.truept\leftarrow\kern-5.truept\to\kern-1.truept}\ }

\def\pagina{\vfill\eject}\let\ciao=\bye

\def\st{\scriptscriptstyle}
\def\*{\vskip0.3truecm}

\def\lis#1{{\overline #1}}\def\eg{\hbox{\it e.g.\ }}
\def\ap{\hbox{\it a priori\ }}
\def\ie{\hbox{\it i.e.\ }}

\def\fiat{{}}
\def\\{\hfill\break} \def\={{ \; \equiv \; }}
\def\Im{{\rm\,Im\,}} \def\sign{{\rm
sign\,}}\def\atan{{\,\rm arctg\,}}

\ifnum\aux=1\BOZZA\else\relax\fi
\ifnum\tipoformule=1\let\Eq=\eqno\def\eq{}\let\Eqa=\eqno\def\eqa{}
\def\equ{{}}\fi
\def\defi{\,{\buildrel def \over =}\,}

\def\1{\ifnum\mgnf=0\pagina\else\relax\fi}
\def\W#1{#1_{\kern-3pt\lower6.6truept\hbox to 1.1truemm
{$\widetilde{}$\hfill}}\kern2pt\,}
\def\Im{{\rm Im}\,}

\def\FINE{
\*
\0{\it Internet:
Authors' preprints downloadable (latest version) at:

\centerline{\tt http://chimera.roma1.infn.it}
\centerline{\tt http://www.math.rutgers.edu/$\sim$giovanni}

\0Mathematical Physics Preprints (mirror) pages.\\
\sl e-mail: giovanni@ipparco.roma1.infn.it,
gentileg@ipparco.roma1.infn.it,\\
vieri@@ipparco.roma1.infn.it
}}

\def\II{{\cal I}}
\def\NN{{\cal N}}

\def\II{{\cal I}}
\def\NN{{\cal N}}

\def\MM{{\cal M}}
\def\AA{{\V A}}

\def\IG{{\cal I}}\def\OO{{\cal O}}

\def\atan{{\,{\rm arctg}\,}}
\def\aa{{\V \a}}\def\nn{{\V\n}}
\def\pps{{\V \ps}}\def\giu{{\downarrow}}\def\su{{\uparrow}}
\def\oo{{\V \o}}\def\nn{{\V\n}}

\fiat


\centerline{\titolo Pendulum: separatrix splitting.}
\*\*

\centerline{\bf G. Gallavotti, G. Gentile, V. Mastropietro}
\*
\centerline{Universit\`a di Roma 1,2,3 }
\centerline{\Di}

\vskip1.truecm {\bf Abstract:} {\it An exact expression for the
determinant of the splitting matrix is derived: it allows us to
analyze the asymptotic behaviour needed to amend the large angles
theorem proposed in {\rm Ann. Inst. H. Poincar\'e, B-60, 1, 1994}. The
asymtotic validity of Melnokov's formulae is proved for the class of
models considered, which include polynomial perturbations.}
\*\*

\0{\bf\S1. Introduction.}
\numsec=1\numfor=1\*

Recently V. Gelfreich noted that a ``theorem'' in [CG] contains an
error. The theorem gave a lower bound on the splitting angles in a
three degrees of freedom system and it was needed to show the
existence of heteroclinic chains in a class of hamiltonian systems
with the aim of an application to a celestial mechanics problem.  We
correct it here by providing the correct lower bound and, at the same
time, exposing again and in a more meditated form some ideas of [CG].
In the present paper we do not discuss the existence of Arnold's
diffusion, [A2], in our systems. We do not discuss the celestial
mechanics application of [CG] either, as two parts of it (see below)
relied on the erroneous statement, and more work is needed. The
present paper is, therefore, {\it not} a correction of the
implications of the error in [CG] but only of the error itself. In
order to derive the same implications further work is necessary as the
erroneous result was used several times in the last three sections of
[CG]. Each use has therefore to be treated separately.

Our analysis deals mainly with systems with {\it three} different time
scales with the ratio between the smallest to the largest being
$\ll1$: \eg the largest is $\h^{-1/2}$, the intermediate is $1$ and
the smallest is $\h^a$, $a\ge0$ (the $a=0$ case being a limiting two
scales problem). The result will be called the {\it large angles
theorem}, (theorem 2 in \S5, proved in
\S6,\S7 for the systems defined in \equ(2.1),\equ(7.1) below), and
describes a property of a pendulum subject to both a slow periodic
force and to a rapid periodic force (incommensurate to the former)
that we imagine generated by a pair of rotators, whose positions are
given by two angles $\a,\l$ while the pendulum position is given by an
angle $\f$. The angles $\a,\l$ will be respectively called {\it slow}
and {\it fast}. 

When the system is perturbed some of the quasi periodic motions
performed by rotators (or clocks) while the pendulum is in its
unstable equilibrium persist. They become motions in which the
pendulum is slightly moving quasi periodically around its unstable
position (``without ever falling down''): one has therefore an
invariant $2$--dimensional torus in phase space. For each position
$(\a,\l)$ of the rotators there is well defined value of the pendulum
variables $(I,\f)$ as well as of the rotators actions $(A,B)$. And
$(\a,\l)$ evolve quasi--periodically (\ie $(\a,\l)$ can be written as
$\a=\ps_1+ \D_1(\ps_1,\ps_2),\l=\ps_2+\D_2(\ps_1,\ps_2)$ and the time
evolution is simply linear $\pps\to\pps+\oo t$ for a suitable $\oo$).

Obviously the above quasi periodic motion  is unstable and
perturbations will make the pendulum ``fall''. The $2$--dimensional
invariant tori have therefore stable and unstable manifolds: which
coincide in absence of perturbations (because they correpond to the
separatrix motions and the pendulum separatrix is degenerate). However
under perturbations they ``split'' (\ie become distinct $3$
dimensional surfaces). Often, simply by symmetry, there is one
trajectory that lies on both manifolds. Physically it corresponds to a
motion that consists of just one swing of the pendulum from nearby the
unstable position back to it.

If we observe the swinging trajectory at the moment it passes through
the stable equilibrium position, $\f=\p$, we see just a point that can
be identified by the value $\aa_0$, at that moment, of the $\aa$
coordinates (where the two manifolds meet). If we move away from that
point the $\AA$ coordinates on the two manifolds become different, at
the same $\aa$, and their difference is a ``splitting'' vector $\V
Q(\aa)$. The $2\times2$ matrix $D=\dpr_\aa\V Q(\aa_0)$ is called the
{\it intersection matrix}.

The homoclinic angles can be defined as the angles whose tangents are
the eigenvalues of the intersection matrix. The {\it splitting} is
usually defined as $\det D$. When the perturbing frequencies are held
fixed and the perturbation is sent to $0$ there is a well known
asymptotic expression for the splitting, called {\it Melnikov's
formula} and coinciding with the ``first order perturbation theory
result''.

The splitting problem is to find under which conditions {\it
Melnikov's formula} holds when {\it both} the perturbation and the
shortest forcing period are sent to $0$.

Informally our result is (see theorem 2 in \S5 for a formal
statement):
\*

\0{\it Suppose that the forcing frequency is very large, say
$\h^{-1/2}$ times the pendulum frequency, with $\h$ very small.
Suppose also that the second characteristic frequency is relatively
very small, say $\h^a$ times that of the pendulum with $a\ge0$. Then
there are perturbations of size $\e=O(\h^c)$ (for some $c>0$) such
that the separatrix splitting has size given asymptotically by {\it
Melnikov's formula} provided the frequencies of the motion verify a
suitable diophantine condition.  In fact this property holds
generically, under the same conditions, for
perturbations of size $\e=O(\h^c)$ which are trigonometric polynomials.}
\*

There are many examples of systems for which the above property does
not hold, see [S], [DJGS], [RW]. But the forcing frequencies relations are
different.

The correct lower bound estimate on the splitting makes an analysis
{\it to all orders} necessary: the analysis, performed here, becomes
``marginal'' but the final result on the existence of homoclinic
splitting (\ie a lower bound on it) remains valid. Therefore the
present paper corrects the error in \S10 of [CG] as far as its
implications on the size of the splitting are concerned.

The techniques we use here {\it to bypass perturbation theory} were
started in the Appendix A13 of [CG]: they were not pushed too far
because the error made any developments unnecessary for the purposes
of [CG].

The techniques were subsequently developed in [G3] (which {\it does
not} repeat the computational error). The methods of [G3] were not
developed to treat three time scales problems: the aim there being to
prove the smallness of the splitting in {\it two time scales} problems
(\ie in systems not considered here with all rotators with comparably
large velocity, $a=-\fra12$). But they can be {\it easily} extended to
three time scales problems and even lead to a remarkable {\it
nonperturbative and exact} computation of the leading order
(exponentially small) of the intersection matrix, see
\equ(6.12),\equ(7.19): V. Gelfreich stressed necessity of a
nonperturbative analysis by a simple cogent argument.

The reason why we study three time scales systems is simply that they
arise in a celestial mechanics problem of interest to us, see \S12 of
[CG]. But most of our results hold also if the slow time scale is the
same as that of the pendulum, \ie the parameter $a$ above (see also
\equ(2.1)) is $a=0$, see also comments in \S9: this is also an 
interesting case and it has been considered, to some extent, in [RW].

The recent work of [RW] considerably overlaps with ours: even when
$a=0$ (a two time scales problem) it cannot be used to achieve all our
results because our assumptions can violate eq. (15) of [RW] (\ie the
bound on the constant called $b$ in [RW], see \equ(2.2), below, for
the isochronous case and \equ(7.3) for the anisochrounus). We also
require, as an essential assumption, that the perturbations be
trigonometric polynomials while the examples of [RW] require, as an
essential assumption, that the perturbation has infinitely many non
vanishing modes (see eq. (19) of [RW]). Another essential difference
with respect to [G3] and [RW] is that we {\it only} study the
splitting {\it at the homoclinic point} while they study it
everywhere: hence our work is much more limited in scope.  On the
other hand we recover some of the results of [RW] because our proofs
also apply nontrivially to two time scale cases with $3$ degrees of
freedom (the ones in [RW]). Nevertheless we feel that the main
difference between our work and [RW] lies in the techniques: here we
show that the techniques of [G3] do apply immediately to the
problem. See concluding remarks for a more technical comparison.
\*
\0{\bf\S2. Isochronous clock--pendulum system.}
\numsec=2\numfor=1\*

Our main result concerns anisochronous systems, \equ(7.1).
Isochronous models will be considered only to illustrate the simplest
cases: the cancellations that we find in the anisochronous case would
look miraculous otherwise.  Calling $(I,\f),(A,\a),(B,\l)$ pairs of
canonical coordinates suppose them action--angle variables:
$(\f,\a,\l)$ is a triple of angles (varying on the $3$--dimensional
torus $T^3=[0,2\p]^3$) and $(I,A,B)\in R^3$.

The hamiltonian will be:

$$H= \h^a A+\fra {I^2}{2}+\h^{-1/2}B+\,g^2\, (\cos \f -1)
+ \m\, \h^c\,f(\f,\a,\l),\ \e\defi\m\h^c\Eq(2.1)$$
where $\h,\m>0$ and $f$ is an {\it even trigonometric
polynomials} in the angles $(\f,\a,\l)$, \ie for instance one could
take $f(\f,\a,\l)=\big(\cos (\a+\f)+\cos(\l+\f)\big)$. The
parameter $\h$ sets the ratios between time scales and is a free
parameter which we take, eventually, to be close to zero in
order to study asymptotic properties as $\h\to0$.

If $\m=0$ the $2$--dimensional torus: $A=B=I=0,\,\f=0$ and
$\aa=(\a,\l)$ arbitrary, is invariant and run quasi-periodically with
rotation velocity $\oo=(\h^a,\h^{-1/2})$: it will be called the {\it
unperturbed torus}.

In the above case, {\it \ie in the isochronous case only}, we shall
suppose (not ``for simplicity'': but as an essential hypothesis) that
the vector $\V\o=(\h^a,\h^{-1/2})$ is a {\it diophantine vector}:

$$|\V \o\cdot\nn|> C^{-1}\h^d{|\nn|^{-\t}}\Eq(2.2)$$
for some {\it diophantine constant} $C,\t$ and some $d>0$. This
restricts the values of $\h$ that we can consider. For $a\in
[0,\fra12]$ it still allows sequences $\h_k$, such that \equ(2.2)
holds with prefixed $\t>2, d=a$ and $C$ large enough, with
$\h=\h_k=\h_1 k^{-1}$, for some $\h_1\in [\fra12,1]$ and all integers
$k$ large enough.
\*

\0{\it Remarks:} 
(1) A related $H$ with $a=\fra12$ arises in a celestial mechanics
problem near a {\it double resonance}, responsible for the time scales
differences (see (12.39) in [CG], after scaling away the factors
$\o_T$ to put $H$ in dimensionless form and with several factors
replaced here by constants, for simplicity). A first simplification of
\equ(2.1) compared to the ``realistic'' model in
\S12 of [CG] is the absence of an additive isochrony breaking term
$\h\fra{A^2}2$: taking it into account does not essentially change the
analysis of the splitting results (even in its quantitative aspects on
the asymptotics as $\h\to0$, see \S7). A second simplification is the
absence in \equ(2.1) of a further perturbation $\b f_0$ which is not
small (\ie $f_0,\b$ are $\h,\m$--independent) but depends only on the
``fast'' angle $\l$ and on $\f$: $f_0=f_0(\f,\l)$. Taking it into
account is a problem not discussed here. However it does not change
the qualitative aspects but only the quantitative ones {\it as long as
the system is isochronous}, see the discussion at the end of \S8.

(2) $g^2$ in the above hamiltonian is {\it fixed} (\ie it is $\h,\m$
independent): eventually we take $g\=1$ for simplicity.  The
parameters $\m,\h$ are free and we shall be interested in them having
a ``small value'': note that if $\h\to0$ the rotation vector $\oo$ of
the unperturbed torus ($I=\f=0$, $B=A=0$) has a size tending to
$\io$. The even parity assumptions on the ``interactions'' $f$
simplifies, possibly in an essential way, the analysis.

In the precession problem of [CG] $\h$ is the deviation
from spherical shape of a planet precessing around its baricenter,
which moves on an ellipse of (fixed) eccentricity $\e=\m\h^c$.

(3) A general physical interpretation of \equ(2.1) is that of a system
consisting in (i) a forcing clock with angular velocity $\h^a$ (\ie a
point moving on a circle with angular velocity $\h^a$, position
$\a$ and action variable $A$); (ii) a pendulum (\ie a point moving on
a vertically placed circle with angular momentum $I$ and position $\f$
counted by taking $\f=0$ as the unstable equilibrium position of the
pendulum); (iii) a second forcing clock with angular velocity
$\h^{-1/2}$ and action variable $B$. Equivalently one can delete the
$A\h^a,B\h^{-1/2}$ terms and replace $\a,\l$ by $\oo t=(\h^a
t,\h^{-1/2} t)$ thus regarding the system as a time dependent one,
consisting of a pendulum subject to a quasi periodic force with
periods $2\p\h^{-a}, 2\p\h^{1/2}$. The ``characteristic time'' of the
pendulum system is $T_0=g^{-1}$.

(4) Finally the ``coupling constant'' in \equ(2.1) is written as $\m
\h^c$ and not just $\m$, because the convergence radius of the 
expansions around the unperturbed torus is expected to be of the order
of some power of $\h$: it would be nice to know the best value of $c$
(thus replacing the constant $c$ by its optimal value) but it seems
not known, see [HMS], [ACKR] (it seems that $c>\fra12$ might be the
right condition, but below we make no attempt at getting even close to
such a small value).
\*

\0{\bf\S3. Separatrices and non degeneracy.}
\numsec=3\numfor=1\*

Supposing $\m=0$ in \equ(2.1) we look at the unstable quasi
periodic motion with $A=0, B=0, I=0,\f=0 $ where $\a=\a_0+{\h}^{a}t$
and $\l=\l_0+\h^{-1/2} t$. This is a family of motions whose initial
data are parameterized by $\V\a\defi(\a_0,\l_0)$ and therefore form an
invariant torus: we shall call it a {\it hyperbolic torus}.  The
energy of such motions is $H=0$. If we fix $A,B$ at other values we
find a continuum of invariant hypebolic tori, with rotation vector
$\oo\defi (\h^{a},\h^{-1/2})$. The energy of such motions is $
\h^{-1/2} B+\h^a A$.

The torus is unstable and its unstable manifold $W^-$ is
$3$--dimensional and contains the set parameterized by $(\f,\a,\l)$ via
the equations:

$$A=0, \qquad I=I(\f), \qquad (\a,\l)\in [0,2\p]^2,\,
\f\in(0,2\p)\Eq(3.1)$$
where $I(\f)=I^0(\f)\defi g\sqrt{2(1-\cos\f)} $ is the pendulum {\it
separatrix}. But also the set with $I=-I^0(\f)$ is part of the torus
unstable manifold because of the meaning of $\f$ as an angle.  The
stable manifold $W^+$ has equations $I(\f)=\pm I^0(\f)$: this means
that $W^+\=W^-$: a well known degeneracy of the pendulum motion.

A nice way of seeing the separatrix motions is by representing them in
the (canonical) {\it Jacobi's coordinates}: these are coordinates $(p,q)$
in terms of which the neighborhood of the (two) lines $I=\pm I^0(\f)$ in
\equ(3.1) is represented, near $I=\f=0$, as:

$$I= R_0(p,q),\qquad \f= S_0(p,q)\Eq(3.2)$$
where the functions $R_0,S_0$ are defined near the origin, where they
vanish, and the pendulum motion becomes, in such coordinates, simply
$p(t)=p_0 e^{-gt}, q(t)=q_0 e^{gt}$.  Therefore the $(p,q)$
coordinates describe globally, as $t\to+\io$, the evolution of initial
data $(p_0,0)$ and globally, as $t\to-\io$, those with data $(0,q_0)$.
In such coordinates the equation of the stable manifold is
simply $p=0$ while that of the unstable manifold is $q=0$.

The functions $S_0,R_0$ have well known holomorphy properties: the
latter imply that the singularities of $R_0(pe^{gt},0),
R_0(0,qe^{-gt})$, at fixed $p$ or $q$, occur at $t=\pm i\fra\p2
g^{-1}$, and the same holds for $S_0$. This can be seen from the
explicit solution of the pendulum equation in terms of elliptic
integrals: an elementary analysis is in Appendix A9 of [CG], or
in Appendix A1 of [G5].

The first result that we use to set up a general picture but strictly
speaking not even necessary as shown in [Ge1], is the (well known, see
[Gr]) stability of such unperturbed hyperbolic torus and of its stable
and unstable manifolds. This will be stated as: \*

\0{\bf Theorem 1}: {\it Suppose  $c$ in \equ(2.1) large enough,
and suppose that $\h$ is such that the rotation vector
$\V\o\defi(\h^a,\h^{-1/2})$ verifies the diophantine property
\equ(2.2) and $\m_0>0$ is small enough. One can define functions
$\G,\X,\L,\Th$ divisible by $\m$ and analytic in the variables
$(\psi_1,\psi_2, p,q,\m)$, varying respectively on the torus
$[0,2\p]^2$, on a neighborhood of $p=q=0$, and in $|\m|\le\m_0$ with
$\m_0$ small enough, such that setting, as $\V\ps$ varies on
$T^2=[0,2\p]^2$:
$$\eqalign{ A=&A_0+ \X(\V\ps,p,q),\quad B=B_0+\G(\V\ps,p,q),\quad
\a=\psi_1,\quad \l=\ps_2\cr
I=& R_0(p,q) +\L(\V\ps,p,q),\quad\f=S_0(p,q)+
\Th(\V\ps,p,q)\cr}\Eq(3.3)$$
one defines, for all $A_0,B_0$, an invariant set on which the motion
described by \equ(2.1) takes place following $t\to (\V\psi+\V\o t, p
e^{-\lis g t}, q e^{+\lis g t})$ (as long as the $(p,q)$ stay in the
domain of definition of $\X,\G,\L,\Th$), with $\lis g=(1+\g(\m))g$ and
$\g$ analytic in $\m$, near $\m=0$ and divisible by $\m$. The
functions $\G,\X$ evaluated at $p=q=0$ have zero average with respect
to $\V\ps$; also the time average of $H$ vanishes on the above motions
and (therefore) $H=0$ for all of them.

The radius of convergence $\m_0$ is uniform in $\h<1$, as long as $\V
\o$ verifies the diophantine condition.} \*

{\it Remarks:} (1) The theorem implies that if $p=q=0$ then the
parametric equations $\f=\Th(\V\ps,0,0)$, $\a=\psi_1$, $\l=\psi_2$,
$I=\L(\V\ps,0,0)$, $A=\X(\V \ps,0,0)$ and $B=\G(\V \ps,0,0)$ describe,
as $\V\ps$ varies on the $2$--dimensional torus $[0,2\p]^2$, an
invariant torus.  The quasi periodic rotation $\V\ps\to\V\ps+\oo t$
with $\oo=(\h^a,\h^{-1/2})$ gives, for all $\V\ps$, a solution to the
equations of motion.  This means that the invariant torus of dimension
$2$ that we considered in \S2 (\ie $A=B=I=0,\,\f=0$) {\it survives the
onset of perturbation} and persists, {\it slightly deformed}, with the
same rotation vector $\oo$ and a slightly varied pair of Lyapunov
exponents (\ie $\pm\lis g=\pm(1+\g)\,g$ rather than $\pm g$).  The
zero average property for $\G,\X$ means that the perturbed torus is in
the average (over $\V\ps$) located at the same position as the
unperturbed one from which it ``emanates'': this is useful as it
allows us to parameterize the invariant tori by their average position
in action space.

(2) Setting $p=0, q\ne0$ one finds a surface of dimension $3$ which is
{\it a part} of the {\it unstable} manifold $W^-$ of the torus, while
setting $q=0, p\ne0$ one gets a part of the stable manifod $W^+$. Such
manifolds are colorfully called local {\it whiskers}. They are in fact
``local'' parts of larger ``global'' manifolds (see item (4) below)
and can be called, for this reason, {\it local stable and local unstable
manifolds}.

(3) The motion of $A,B$ is computed from that of the other coordinates
by quadrature (\eg $\dot B=-\e \,\dpr_\l f(\f,\a,\l)$, while $B$
itself does not occurr in the equations of the other coordinates, and
similarly for $A$).  The symmetry of $f$ does not imply that $\L,\Th$
have zero average over $\V \ps$, if $\f=\p$, [CG], and also the
Lyapunov exponent $\lis g$ in general changes by a quantity $\g$ of
order $O(\m)$ with respect to the unperturbed value $g(pq)$. The
motions energy also changes, in general, by an amount analytic in
$\m$ and divisible by $\m$ (\ie of $O(\m)$).

(4) Once the {\it local} parts of the torus whiskers have been defined
as above we can extend them to {\it global} objects by simply applying
time evolution to their points.

(5) The case $\m=0$ (one of the most classical results in Mechanics)
is a non trivial exercise: it is developed in Appendix A9 of [CG] or
Appendix A1 of [G5].

(6) The stable and unstable manifolds do not coincide, in general, for
$\m\ne0$.

(7) in the anisochronous cases of \S7 a very similar result holds with
the difference that the relation between $\aa$ and $\pps$ is also non
trivial and described by a function $\V \D(\pps,p,q)=(\D(\pps,p,q),0)$
with zero average over $\pps$ and divisible by $\m$ and with the same
domain of definition of the other functions in \equ(3.3). In this case
all the functions in \equ(3.3), and $\V\D$ as well, depend also on
$A_0$ which must be restricted so that $\oo=(\h^a+\h A_0,\h^{-1/2})$
verifies a diophantine property with suitable diophantine constants
$C(\h),\t$: the size of the radius of convergence and the value of $c$
will depend on the selected $C(\h)$ (like \equ(2.2) or \equ(7.3)).

(8) The above theorem is well known; it is explicitly proved (in a
much more general case) in the above form in \S5 of [CG], see
p. 38. Its proof is a rather straightforward adaptation of Arnold's
method of proof of the KAM theorem, [A1], as exposed for instance in
[G1]: the only element of ``novelty'' is perhaps the {\it normal form}
of the motion in the coordinates $\V\ps,p,q$, as long as the latter
two remain in their domain of definition.  A more modern proof, based
on Eliasson's method, [E], for proving KAM (as exposed in [G2],[GG])
and extending it to the problem of tori of one dimension less than the
maximal, can be derived from [Ge1] (where only the cases $p=0$ or
$q=0$ are studied).  Also unusual is the absence of the twist
condition (present, and necessary, in the more general proof in [CG]):
it can be eliminated because of the special structure of the
hamiltonian \equ(2.1). But it is a long and uninteresting proof for us
here.  We shall not need, however, the normal form and, as it will be
clear, we only need the classical results in the weaker form discussed
in [Gr],[Ge1].
\*

\0{\bf\S4. Splitting angles. A recursive determination.}
\numsec=4\numfor=1\*

The symmetry of $f$ implies an intersection at $\f=\p$ and $\V\a=\V0$
(see below or, for instance, p. 363 of [G3]) between stable and
unstable manifolds of the torus into which the unperturbed torus
(\ie $A=B=0,I=0,f=0$) is deformed by the perturbation.  Therefore we set
up an algorithm to study such intersection. For this purpose it is
convenient to work in the original canonical coordinates and write the
stable and unstable whiskers $W^\pm_\m$ as:

$$W^\pm_\m=\{(\f,\aa,I,\AA)=\big(\f,\aa,
I^\pm_\m(\f,\aa),\AA^\pm_\m(\f,\aa)\big)\}\Eq(4.1)$$
with $ \aa\in T^2,\,\e< |\f|<2\p-\e$ where $\e>0$ can be fixed \ap as
small as we please provided we diminish the value of the analyticity
radius $\m_0$ in theorem 1.

In other words the whiskers deformation is of $O(\m)$ in every closed
subinterval of $(-2\p,2\p)$: therefore they remain
parametrizable by $\f,\aa$ for $\f$ in any closed subinterval if $\m$ is
small enough (just note that for $\m=0$ they {\it are}
parametrizable). We say that, in such region of $(\f,\aa)$, they
are {\it graphs} over the angle variables.  We only need that $\f=\p$ is
allowed (hence $\e=\fra\p2$ will do).

We define the {\it splitting vector} $\V Q(\aa)$, the {\it splitting }
or {\it intersection matrix} and the {\it splitting} between
$W^+_\m$ and $W^-_\m$ at $\f=\p$ and $\V\a=\V0$, respectively, as:

$$\V Q(\aa)\defi \AA^+_\m(\p,\aa)-\AA^-_\m(\p,\aa),\
D\defi \dpr_\aa \V Q(\aa)\big|_{\aa=\V0},\ {\rm and}\ \det D\Eq(4.2)$$
A relevant remark (Lochak and Sauzin, private communication) is that
the whiskers are {\it lagrangian manifolds} so that, for a suitable
generating function $S^\pm(\f,\aa)$, the whiskers have equation:
$\AA^\pm=\dpr_\aa S(\f,\aa), I^\pm=\dpr_\f S^\pm(\f,\aa)$ around every
point $(\f_0,\aa_0)$ where they can be locally regarded as graphs over
the angles $(\f,\aa)$, \eg at $\f=\p$ and for $\m$ small. This implies
that $D$ is {\it symmetric}, as we indeed find in \equ(6.10),
\equ(7.10).

We now derive recursive formulae for $I^\pm_\m$, $\AA^\pm_\m$ in
\equ(4.1) and their time evolution, keeping in mind that for $\m=0$ it
is $I^+_0=I^-_0=I^0(\f)=g \sqrt{2(1-\cos\f)}, \AA^\pm_0=\V0$.  The
unperturbed motion is simply: $X^0(t)\=(\f^0(t),
\aa+\oo t,I^0(\f^0(t),\V 0)$, where $(\f^0(t),I^0(\f^0(t)))$ is the free
(\ie with $\m=0$) separatrix motion, generated by the pendulum in
\equ(2.1) setting the origin of time when the pendulum swings through
$\f=\p$. The following \S4,\S5 are presented here only for
completeness: although selfcontained {\it they are not} meant as a
substitute of the work done in [G3] but serve the purpose of guiding
the reader to dig out of that paper what he may want to see in more
detail.

Let $X_\m^\s(t;\aa)$, $\s=\pm$, be the evolution of the point on
$W^\s_\m$ whose initial coordinates are given by
$(\p,\aa,I^\s_\m(\p,\aa),\AA^\s_\m(\p,\aa))$; from now on we shall fix
initial data with $\f=\p$ (which amounts to studying the whiskers at
the ``Poincar\'e's section" $\{\f=\p\}$).

The analyticity in $\m$ implied by the above theorem 1 and the
analyticity properties of the Jacobi functions $R_0,S_0$ allow us to
consider the {\it convergent} (if $c$ in \equ(2.1) theorem 1 is large)
Taylor series expansions, in $\m$, of the whiskers equations. Let:
$$X^\s_\m(t)\=X^\s_\m(t;\aa)\defi \sum_{k\ge 0} X^{k\s}(t;\aa) \e^k\=
\sum_{k\ge 0} X^{k\s}(t) \e^k,\qquad \s=\pm\Eq(4.3)$$
be the power series in $\e=\m \h^c$ of $X^\s_\m$, (convergent for $\m$
small); note that $X^{0\s}\=X^0$ is the degenerate unperturbed
whisker. We shall often omit writing explicitly the $\aa$ variable
among the arguments of various $\aa$ dependent functions, to simplify
the notations.

Theorem 1, \S3, tells us that the $t$--dependence of $X^\s_\m(t)$ has
the form:
$$X^\s_\m(t)=X^\s_\m(\oo t,t;\aa)\Eq(4.4)$$
where $X^\s_\m(\V\ps,t;\aa)$ is a real analytic function, of {\it all}
its arguments ($\m$ included), which is periodic in $\V\ps$ and
$\aa$. And in fact in our {\it isochronous} case the dependence on $\V\ps$
and $\aa$ is via $\aa+\V\ps$ (``no phase shift'' in the sense of
[CG]). The two functions $X^{k\s}(t)$ will be regarded as forming a
single function $X^k(t)$:
$$X^k(t)=\cases{X^{k+}(t)& if $\s={\rm sign}\,t=+$\cr
X^{k-}(t)& if $\s={\rm sign}\,t =-$\cr}\Eq(4.5)$$
{\it We label the $6$ components of $X$ with an index $j$,
$j=0,1,\ldots,5$, and write them ({\rm notation used in [G3]}),
with the convention:
$$X_0\defi X_-,\quad (X_j)_{j=1,2}\defi \V X_\giu,\quad
X_3\defi X_+,\quad (X_j)_{j=4,5}\defi \V X_\su\Eq(4.6)$$
and we write first the angle variables ($(\f,\a,\l)=(X_-,X_\giu)$),
then the action variables ($(I,A,B)=(X_+,\V X_\su)$); first the pendulum,
then the rotator and then the clock variables.}

Therefore at order $0$ in $\m$:
$$\eqalign{
X_0(t)=&\f^0(t),\quad X_1(t)=\a_0+\h^a t,
\quad X_2(t)=\l_0+ \h^{-1/2}t\cr
X_3(t)=& I^0(\f^0(t)),\quad X_4=0=X_5,\cr }\Eq(4.7)$$
where $\f^0(t)$ is the free separatrix motion (\ie $\f^0(t)=4\atan e^{-g
t}$, see Appendix A1).

Inserting \equ(4.3) in Hamilton's equations for \equ(2.1) and comparing
the various orders in $\m$, the coefficients $X^{k\s}(t)\=$\-
$X^{k\s}(\oo t,t;\aa)$ are seen to satisfy the hierarchy of equations:
$${d\over dt} X^{k\s}\= \dot {X}^{k\s}=L X^{k\s}+F^{k\s}\Eq(4.8)$$
where $L$ is a $6\times 6$ matrix with only two non vanishing elements
$L_{03}=1$ and $L_{30}=g^2\cos\f^0(t)$.  Expanding $X^\s$ in powers of
$\e$ and imposing that the equations of motion are verified, we find
recursively the expressions for $F$. For instance:
$$\eqalignno{
F^1_+=&-\dpr_\f f,\qquad \V F^1_\su=-\dpr_\aa f\cr
F^2_+=&-{g^2\over2}\sin\f\,
(X^1_-)^2-\dpr_{\f^2} f X^1_- -\dpr_{\f\aa} f \V X^1_\giu\cr
\V F^2_\su=&-\dpr_{\aa\f}f X^1_--\dpr_{\aa^2}f \V X^1_\giu&\eq(4.9)\cr
F^3_+=&-{g^2\over2}\,2X_-^1X_-^2\sin\f-{g^2\over
3!}\cos\f(X^1_-)^3-\dpr_{\f^2}f\,X_-^2+\cr
&-\dpr_{\aa\f} f
\V X^2_\giu-{1\over2}\dpr_{\f\aa^2}f \V X^1_\giu \V X^1_\giu-\dpr_{\f^2\aa}f
\V X^1_\giu X^1_--{1\over2}\dpr_{\f^3}f\,(X^1_-)^2\cr
\V F^3_\su=&-\dpr_{\aa\f}fX^2_--\dpr_{\aa^2}f
\V X_\giu^2-{1\over2}\dpr_{\aa^3}f \V X^1_\giu
\V X^1_\giu-\dpr_{\f\aa^2}f\,\V X^1_\giu X^1_-- {1\over2}\dpr_{\aa\f^2}f
X^1_-X^1_-\cr}
$$
where the functions are evaluated at $\f(t)=\f^0(t),\aa(t)\defi
\aa+\oo t$. More generally, $F^{k}$ depend upon $X^0,...,X^{k-1}$
{\it but not on $X^{k}$}. It is $F^k_-=\V F^k_\giu\=0$ for all
$k\ge1$.

Expressing the solution of a linear inhomogeneous equation like the
one in \equ(4.8) can be conveniently done in terms of the {\it
wronskian matrix}.  We recall therefore the notion of {\it wronskian
matrix} $W(t)$ of a  solution $t\to x(t)$ of a differential equation
$\dot x= f(x)$ in $R^n$.

It is a $n\times n$ matrix whose columns are formed by $n$ linearly
independent solutions of the linear differential equation obtained by
linearizing $f$ around the solution $x$ and assuming $W(0)=$ identity.
In other words $W(t)$ is, in our case, the solution of the
differential equation $\dot W(t)=L(t)\,W(t),\, W(0)=1$.

The solubility by elementary quadrature of the free pendulum equations
leads, on the separatrix, to the following formulae that have
importance because the wronskian of the free separatrix motion can be
expressed in terms of them. If $\f^0(t)=4 \arctan e^{-gt}$:
$$\eqalign{
\cos \f^0=&1- {2\over (\cosh gt)^2}= 1- 8 {x^2\over (1+x^2)^2}\cr
\sin \f^0=& 2 {\sinh gt\over (\cosh gt)^2} = 4\s x {1-x^2\over
(1+x^2)^2}\cr}\Eq(4.10)$$
for $x=e^{gt}$ {\it or, as well, for} $x=e^{-gt}$.  Eq. \equ(4.10)
leads, see \equ(A1.1) in Appendix A1, to the following expression for
the wronskian $W(t)$ of the {\it separatrix motion} for the pendulum
appearing in \equ(2.1), with initial data at $t=0$ given by
$\f=\p,I=2g$:
$$W(t)=\pmatrix{
{1\over\cosh gt}&{{\lis w}\over4}\cr
-{\sinh gt\over\cosh^2 gt}&
(1-{{\lis w}\over4}{\sinh gt\over\cosh^2gt})\cosh gt\cr},
\qquad{\lis w}\={2gt+\sinh 2gt\over\cosh gt}\Eq(4.11)$$
Notationally we follow here [G3] (in [CG] $I,\f$ are exchanged). The
evolution of the $X_\pm$ (\ie $I,\f$) components can be determined by
using the above wronskian:
$$\pmatrix{X^{k\s}_-\cr X^{k\s}_+\cr}=W(t)
\pmatrix{0\cr X^{k\s}_+(0)\cr} +
W(t)\ig_0^t{W\,}^{-1}(\t)\pmatrix{0\cr F^{k\s}_+(\t)\cr}\ d\t
\Eq(4.12)$$
Thus, denoting by $w_{ij}$ ($i,j=0,3$) the entries of $W$, we see
immediately that:
$$\eqalign{\textstyle X^{k\s}_-(t)=&\textstyle
w_{03}(t)X^{k\s}_+(0)+w_{03}(t) \ig^t_0w_{00}(\t)
F^{k\s}_+(\t) d\t
\textstyle-w_{00}(t)\ig^t_0w_{03}(\t)
F^{k\s}_+(\t)\,d\t\cr
\textstyle X^{k\s}_+(t)=&\textstyle w_{33}(t)X^{k\s}_+(0)+w_{33}(t)
\ig^t_0 w_{00}(\t) F^{k\s}_+(\t) d\t
\textstyle-w_{30}(t)\ig^t_0 w_{03}(\t) F^{k\s}_+(\t)\,d\t\cr}\Eq(4.13)$$
Integrating on the separatrix \equ(4.8) for the $\su,\giu$ components
is ``easier'': one can find it directly or, more systematically, by
writing the full $6\times6$--wronskian matrix of the equation
\equ(4.8), which is trivially related to $W(t)$ above; we shall not
write it here: see eq. (4.21) in [G3] for an explicit expression. The
result is:
$$\V X_\giu^{k\s}(t)= \V 0,\qquad
\V X_\su^{k\s}(t)=\V X_\su^{k\s}(0)+\ig_0^t\V
F^{k\s}_\su(\t)d\t\Eq(4.14)$$
having used that the $\V X^{k\s}_\giu(0)\=\V 0$, because the initial
datum is fixed and $\m$--independent.

If \equ(2.1) is modified by adding an isochrony breaking term $\fra\h2
A^2$ the first component of the first of \equ(4.14) becomes:

$$X^{k\s}_1(t)=\h \,\big(t X^{k\s}_1(0)+\ig_0^t (t-\t)
F_1^{k\s}(\t)\,d\t\big)\Eq(4.15)$$
where $X_1^{k\s}$ is the first compnent of $\V X_\su^{k\s}$, see
eq. (2.18) in [G3].

Equations \equ(4.13), \equ(4.14) can be used to find a reasonably
simple algorithm to represent whiskers to all orders $k\ge1$ of the
perturbation expansion. It is very important to keep in mind that the
initial data in \equ(4.13), \equ(4.14) {\it are not constants}:
according to the convention following \equ(4.3) they can depend on the
$\aa$ variables of the initial data. This means that the functions $X$
depend separately on $\aa$ and $\oo t$.  {\it Except} when, as in
\equ(2.1), the hamiltonian is linear in the $A,B$ variables. In the
latter case the dependence on $\aa$ and $\oo t$ of the r.h.s. of \equ(4.4)
(where the notation is complete and all variables are indicated
explicitly) {\it must be} through $\aa+\oo t$, since the $\aa$ angles
vary as $\dot{\aa}=\oo$.

Note that the case \equ(2.1) is {\it non trivial and, in fact, very
interesting: it is equivalent to a problem on a non linear quasi
periodic Schr\"odinger equation}, see [G2], [BGGM].  The extension to
anisochronus cases (\ie with a quadratic term in $A$ added to
\equ(2.1)) is worked out in [CG] up to third order: a general
analysis can be found in [G3].

The initial data (still unknown) in \equ(4.13),\equ(4.14) are
determined by imposing the correct behavior at $\pm\io$, and the
correct dependence on $\aa$ and $\oo t$ (\ie a dependence on these two
vectors through their sum). These conditions arise from the fact that
the motion must be asymptotic to the quasi periodic motion on the
invariant torus whose whiskers are described by $X^+$ or $X^-$. The
scheme to do so is the following, see [G3].

Note that $w_{03},w_{33}$ in \equ(4.11) behave, as $t\to\s\io$ with
$\s=\pm1$, asymptotically as $\s e^{gt\s}/4$ and $-e^{gt\s}/4$, while
the other two matrix elements become exponentially small. Hence we see
that the tems in \equ(4.13) proportional to $w_{33}(t)$ or $w_{30}(t)$
diverge, in general, as $t\to\s\io$ exponentially fast (supposing the
integrals convergent, \ie supposing $F^{k\s}_+(\t)$ not growing
faster than a polynomial in $t$, as it will turn out to be). But they
are multiplied by: $\big(X^{k\s}_+(0)+\ig_0^t w_{00}(\t)
F_+^{k\s}(\t)\,d\t\big)$ so that the condition to determine
$X^{k\s}_+(0)$ is:
$$X^{k\s}_+(0)+\ig_0^{\s\io} w_{00}(\t)
F_+^{k\s}(\t)\,d\t=0\Eq(4.16)$$
Likewise, in the isochronous case, from the second of \equ(4.14) we
determine $\V X^{k\s}_\su(0)$ by imposing that $\V X_\su^{k\s}(t)$,
\ie the momentum component corresponding to the isochronous angles
$\a,\l$, depends asymptotically on $t$ only via $\aa+\oo t$: this will
determine $\V X^{k\s}_\su(0)$ up to a constant.

And the average over $\aa$ of $\V F^{k\s}_\su(t)$ must tend to $0$ as
$t\to\s\io$ (otherwise the second of \equ(4.14) could not possibly be
bounded as $t\to\s\io$: but it has to be such because $\V X_\su^{k\s}(t)$
has to be bounded, by theorem 1).  This means that the constant is not
a function of $\aa$ and can be fixed arbitrarily: however we want that
the averages of the $\Xi,\G$ in \equ(3.3) are $0$ so that $X^{k\s}(0)$
is completely determined. One finds, for {\it both components}:

$$\V X^{k\s}_\su(0)+\ig_0^{\s\io} \V F^{k\s}_{\su}(\t) d\t=\V0\Eq(4.17)$$
where the integral is usually improper: see \equ(5.1) below for a
proper definition (derived by simply looking at the meaning of the
conditions imposed to determine $\V X^{k\s}_\su(0)$). Eq. \equ(4.17)
{\it remains the same} even in the anisochronous cases (see eq. (4.5)
in [G3]).

The key to concrete calculations is that, $f$ in \equ(2.1) being a
trigonometric polynomial, the function $F^1$ (see \equ(4.9)) belongs
to the class $\MM$ of linear combinations of terms like:
$$M(t)={(g\s t)^h\over h!} x^k \s^\th e^{i\r\oo\cdot\nn t}\Eq(4.18)$$
with $h\ge 0,k$ integers, $\th=0,1$, $\r=\pm1$ and $gk \pm i\oo\cdot
\nn\ne 0$ and we set $x=e^{-gt\s}$ with $\s=\sign\,t=\pm1$.
In fact $F^1$ is a {\it finite linear combination} of harmonics
$\nn$. By \equ(4.10) we see that \equ(4.11) is an analytic function of
the variable $x=e^{gt}$ and of $t$, or of $x=e^{-gt}$ and of $t$: the
explicit $t$--dependence arises because of the term $\lis w$ in
\equ(4.11).

Expressing $F^1$ as a sum of monomials like \equ(4.18) requires a
(convergent) infinite sum over $k,h$.  In the case of $F^1$, in fact,
there are {\it no monomials} with a power of $t$ higher than $0$ (\ie
with $h>0$ in \equ(4.18)).  By induction {\it all} the $F^k$ have the
property of being expressible as sums of monomials like \equ(4.18).
However, starting with the second order, one sees that powers of
$t$ do appear (this reflects, see theorem 1, that the function $\g$,
describing the change of the Lyapunov exponents to $\pm (1+\g) g$ with
$\g$ analytic in $\m$, is not identically zero; see, however, [Ge2]).
A full description of the induction can be found in [G3].

For the analyticity properties of the series introduced above we refer
to \S2 of [G3] and we proceed to a quick discussion of the
determination of the the initial constants.
\*

\0{\bf\S5. Improper integrals and the operators $\II,\OO,\OO_0$.}
\numsec=5\numfor=1
\*

The integrations in \equ(4.13),\equ(4.14) can be
expressed in terms of an operator $\II$ acting linearly on finite
linear combinations of monomials like \equ(4.18) with
$k^2+(\oo\cdot\nn)^2>0$:
$$\eqalign{
&\II M(t)\defi \ig_{\s\io}^t M(\t) d\t, \qquad {\rm with:}\cr
&\II M(t)=-   \s^{\th +1} x^k e^{i\r\oo\cdot\nn t}
\sum_{p=0}^h{ (\s t)^{h-p} \over(h-p)!}
{1 \over(k- i \r \s  \oo\cdot\nn)^{p+1}}\cr} \Eq(5.1)$$
where the first row is a formal definition whose mathematical meaning
is given by the second row (note that if $k\le0$ the first line is an
improper integral), and we set $g=1$.

Note that the $\II$ is {\it not defined} on the polynomials of $t,\s$,
\ie if $k=0$ and $\oo\cdot\nn=0$ (so that no exponentials are present
in the monomial defining $M$). It can be naturally extended, for
$j\ge0$, to the polynomials by setting $\II
t^j\s^\th=\fra{\s^\th}{j+1} t^{j+1}$, see (3.7) in [G3].

The $\II$ is an integration with respect to $t$
with special initial data: in fact {\it at fixed $\s$}:
$$\dpr_t\II M\=M\Eq(5.2)$$
If $M$ is such that $M(t)\=M(\oo t,\s)$ for some $M(\pps,\s)$
defined on the torus, then:
$$\II M(t)=(\oo\cdot\V\dpr_\pps)^{-1}M(\oo t,\s)\Eq(5.3)$$
The integrals in \equ(4.13),\equ(4.14) can be expressed in terms of
the operators:
$$\eqalign{
\OO F(t)= &w_{03}(t)\bigl(\IG (\,w_{00}
F))\bigr)(t)-w_{00}(t)\bigl(\IG (
w_{03}F)\bigr)(\t)|^t_{0^\s}\cr
\OO_+
F(t)=& w_{33}(t)\bigl(\IG (\,w_{00} F)\bigr)(t)- w_{30}(t)\bigl(\IG
w_{03})F \bigr)(\t)|^t_{0^\s}\cr
\lis\II^2 F(t)=&\II^2 F(t)-\II^2 F(0^\s)\cr
}\Eq(5.4)$$
where $\s=\sign t $. Then one finds, in the general anisochronous
case:
$$X^h_-(t)=\OO F^h_+(t),\quad X^h_+(t)=\OO_+ F^h_+,\quad
\V X^h_\giu(t)=\h E\lis\II^2 F^h_\su(t),
\quad \V X^h_\su(t)=\IG \V F^h_\su(t)\Eq(5.5)$$
where $E$ is the projection over the first component, see \equ(4.15),
and $F^{h}$ have to be expressed in terms of the $X^{h'}$ with
$h'<h$.

Since, in this section, we consider an isochronous model the above
formulae are slightly simpler as $\V X^h_\giu(t)\=\V0$, \ie must take
$E=0$ in the third of \equ(5.5). In \S7 we shall need, however,
\equ(5.5).

There is no difficulty in setting up a general recursive
scheme for the computation. We just give the result using the
convenient notations ($k^i_j\ge 1, m_i>0$):
$$
(k^i_j)_{\V m,p}\=(k^0_1,\ldots,k^0_{m_0},k^1_1,\ldots,k^1_{m_1},
\ldots,k^{3}_1,\ldots,k^{3}_{m_{3}})\qquad {\rm
s.t.\ }\sum k^i_j=p\Eq(5.6)$$
referring to \S2 of [G3] for more details (if needed). If $f_1\defi f,
f_0\defi g^2\cos\f$, we find:
$$\eqalignno{
& F_-^{k\s}\=0\ ,\  \V F_\giu^{k\s}\=\V 0\ ,\
\V F_\su^{k\s} = - \sum_{|\V m|\ge0,\d=0,1}\fra1{\V m!}
(\dpr^{\V m}\dpr_\aa f_\d)
\sum_{(k^i_j)_{\V m,k-1}} \prod_{i=0}^{l-1}\prod_{j=1}^{m_i}
X^{k^i_j\s}_i\cr
& F_+^{k\s} \=- \sum_{|\V m|\ge 2,\d=0,1}^* \fra1{\V m!}
(\dpr^{\V m}\dpr_\f f_\d)
\sum_{|\V m|\ge0} (\dpr_0 f_\d)_{\V m}
\sum_{(k^i_j)_{\V m,k-1}} \prod_{i=0}^{l-1}\prod_{j=1}^{m_i}
X^{k^i_j\s}_i&\eq(5.7)\cr}$$
where $(k^i_j)_{\V m,k},(k^i_j)_{\V m,k-1}$ are defined in
\equ(5.6).  The $*$ means that if $\d=0$ only vectors $\V m$ with
$|\V m|\ge2$ have to be considered in the sums.  Note that if $\d=0$
the sum in the expression for $F^h_+$ can only involve vectors $\V m$
with $m_j=0$ if $j\ge1$, because the function $f_0= g^2\cos\f$ depends
only on $\f$ and not on $\aa$, (hence also $k^i_j=0$ if $i>0$). The
functions are evaluated at $\f(t)=\f^0(t),\aa(t)\defi \aa+\oo t$.
The indices in \equ(5.7) are mutually contracted with a natural rule
that we leave to the reader to work out.

The relations $\V F^h_\giu=\V0$ and $F_-^h\=0$ are general (as the
equations for $\dot {{\V X}}_\giu, \dot X_-$ are linear) and in the
{\it isochronous cases}:
$$\eqalign{
\V X_\su^1(t)=&-\II(\dpr_\aa f),\quad
\V X^2_\su(t)=-\II\Big(\dpr_{\aa\f} f\,\OO(-\dpr_\f f)\Big)
\cr
\V X^3_\su(t)=&-\II\Big(\dpr_{\aa\f} f\OO\big(
-{1\over2}\sin\f\OO(\dpr_\f f)\OO(\dpr_\f f)\big)\Big)+\cr
&-\II\Big(\dpr_{\aa\f} f\OO\big(\dpr_{\f^2} f\OO(\dpr_\f f)\big)\Big)
-{1\over2}\II\Big(\dpr_{\aa\f^2} f\OO(\dpr_\f f)\OO(\dpr_\f
f)\Big)\cr}\Eq(5.8)$$
(see the examples in \equ(4.9)).  We fix our attention on the
models with $f$ a trigonometric polynomial (``trigonometric
perturbation'') of degree $N$:
$$\eqalign{
&f(\aa,\f)=f_S(\f,\a,\l)+f_F(\f,\a,\l)\cr
& f_j(\f,\a,\l)=
\sum_{n,\nn}^N f_{j,(n,\nn)}\,\cos(\n_1\a+\n_2\l+ n\f),\quad
j=S,F\cr}\Eq(5.9)$$
with $|n|,|\nn|\le N$ and $f_{S,(n,\nn)}=0$ unless $\n_2=0$, \ie $\nn$
is a {\it slow} mode, while $f_{F,(n,\nn)}=0$ unless $\nn_2\ne0$, \ie
$\nn$ is a {\it fast} mode. We also say that $f_S$ depends only on
slowly rotating angles and $f_F$ depends on fastly rotating
angles.  A nontrivial example can be:
$$f(\a,\l)=\big(\cos(\a+\f)+\cos(\l+\f)\big)\Eq(5.10)$$
The intersection matrix to order $h$, $D^h_{ij}$, can be expressed,
to order $h=1,2,3$ as:

$$\eqalignno{
D^1_{ij}=&\ig_{-\io}^\io dt \,\dpr_{ij}f,\quad
D^2_{ij}=- \ig_{-\io}^\io dt\,\big[\dpr_{ij0}f\,\OO(\dpr_0 f)+
\dpr_{j0} f \OO(\dpr_{i0} f)\big]\cr
D^3_{22}=&\ig_{-\io}^\io dt\,\Big[
w_{30}\,\OO(\dpr_{220}f)\,\OO(\dpr_0 f)^2+2w_{30}\,\OO(\dpr_{20} f)^2
\,\OO(\dpr_0 f)+&\eq(5.11)\cr
&+ \dpr_{200}f\, \OO(\dpr_{20}f)\,\OO(\dpr_0f)+\dpr_{00}f\,\OO(\dpr_{220}f)
\,\OO(\dpr_0 f)+\cr
& +\dpr_{00}f\,\OO(\dpr_{20}f)\,\OO(\dpr_{20} f)+\fra12
\dpr_{2200} f\,\OO(\dpr_0 f)^2+ \dpr_{200}f
\OO(\dpr_{20}f)\,\OO(\dpr_{0}f)\Big]\cr}$$
where the derivatives of the $f$'s are evaluated at $\f(t),\aa+\oo t$,
with $\f(t)\=\f^0(t)$ the free separatrix motion, see \equ(4.10). We
set $\dpr_0\defi \dpr_\f$ and $\dpr_i\defi \dpr_{\a_i}$;
the $\aa$'s have to be set equal to $\00$ after evaluating
derivatives.  The expressions contain improperly convergent
integrals (in general) and must be understood by thinking
$\ig_{-\io}^{+\io}$ as $\ig_{-\io}^{0}+\ig_{0}^{+\io}$ and by using
the definition \equ(5.1), see [G3].

It is convenient to split the operation $\OO$, see (6.5) in [G3],
as:

$$\eqalign{
\OO(F)=& \OO_0(F)+|w_{03}(t)|\, G^{(0)}(F)+\, w_{00}(t)\,
G^{(1)}(F)\cr
\OO_0(F)=&\fra12\Big(\ig_{-\io}^t d\t (w_{03}(t)\,w_{00}(\t)
-w_{00}(t)\,w_{03}(\t))\,F(\t)+
\ig_{+\io}^t d\t\,(same)\Big)\cr
G^{(0)}(F)=&\fra12\ig_{+\io}^{-\io} d\t\,w_{00}(\t)\,F(\t),
\qquad G^{(1)}(F)=
\fra12\,\ig_{+\io}^{-\io} d\t\,|w_{03}(\t)|\,F(\t)\cr}\Eq(5.12)$$
The identity, see (6.12) and Appendix A2 of [G3]:
$$\ig_{-\io}^{+\io} dt\, F(t)\OO(H)(t)=\ig_{-\io}^{+\io} dt\,
H(t)\OO(F)(t)\Eq(5.13)$$
implies symmetry of the above matrices $D_{ij}$, at least for the
first three orders (see \equ(5.11)): symmetry follows to all
orders as said after \equ(4.2), or as it will be seen in \S6.

In the anisochronous case, \ie in \S7, we shall also use the splitting:

$$\eqalign{
&\lis\II^2(F)=\II^2_0(F)+|t| G^{(2)}(F)+G^{(3)}(F)\cr
&\II^2_0(F)=
\fra12\Big(\ig_{-\io}^t d\t\, (t-\t)\,F(\t)+\ig_{+\io}^t d\t\,(same)\Big)
\cr
&G^{(2)}(F)=\fra12\ig_{+\io}^{-\io}d\t\,F(\t),\qquad G^{(3)}(F)=
\fra12\ig_{+\io}^{-\io}d\t\,|\t|\,F(\t)\cr}\Eq(5.14)$$
see eq. (6.3),(6.6) in [G3].

The key remark, to understand the asymptotic behaviour of \equ(5.11)
as $\h\to0$, is that whenever the integrand is analytic it becomes
possible to shift the integrations, over $t$ and the $\t$'s, to an
axis close to $\Im t$, $\Im\t=\pm(\fra\p2-\h^{1/2})$: see \S8 in [G3]
(choosing the free parameter $\d$ appearing in [G3] as $\h^{1/2}$).

Using that $G^{(1)}(F)=G^{(0)}(F)=0$ if $F$ is odd (as the odd derivatives of
$f$ are, when evaluated at $\aa=\V0$) and using also that $\OO_0$
leaves parity unchanged, in general, we shall find that the {\it non
analytic } terms (\ie those containing integrals of a non analytic
function, \eg $|w_{03}(\t)|$) {\it cancel each other} in their
contribution to the determinant of $D_{ij}$ {\it to all orders}
$k\ge1$. The result will be the proof
of the following theorem:
\*

\noindent{\it {\bf Theorem 2} ({\sl``Large angles theorem''}):
Consider a system described by the hamiltonian \equ(2.1) or \equ(7.1)
with $f$ an even trigonometric polynomial. Let $\m$ be small
($|\m|<\m_0$) and $c$ large, enough.  Consider an invariant torus with
diophantine rotation $\oo$ among those described in theorem 1 above.  At
the homoclinic point with $\f=\p,\aa=\V0$ the intersection matrix
determinant is exponentially small as $\h\to0$ and it is generically
asymptotic to its first order value (``Melnikov' formula'').  The choice
given by
\equ(5.10) is a concrete example of this result, generically holding
for \equ(5.9).}
\*

\noindent{\it Remark}: The result in the case $f$ is given by  
\equ(5.10) is $\det D=24 \p \h^{-1/2} \e^2 e^{-\fra\p2 \h^{-1/2}}$ 
in both cases (isochronous, \equ(2.1), and anisochronous, \equ(7.1))
with $a>0$. This theorem has an analogue with and $a=-\fra12$, \ie
only two time scales: an early review is in [G3]. In the latter case
it has been extended to analytic perturbations, \ie beyond the
trigonometric case, and to cover the exact asymptotic value of the
splitting, \ie far beyond [G3], see [DGJS], [RW].  The above three
scales case, \equ(2.1), is quite different from the two scales case
(discussed in [G3]): but arises naturally in celestial mechanics
problems near a double resonance, as in the case of the precession
problem in [CG] to which we hope to apply, eventually, the results of
this paper.

The case in \equ(2.1) is that of a pair of ``clocks'' and a
pendulum. The case of a ``clock'', a ``rotator'' and a pendulum is
exemplified by the hamiltonian obtained by adding $\fra12 \h A^2$ to
\equ(2.1), see \equ(7.1) below. Both cases are treated in full detail
in \S6,\S7.
\*

\0{\bf\S6. Nonperturbative splitting analysis
in presence of fast and slow rotations.}
\numsec=6\numfor=1\*

It would be easy to show that the determinant of the intersection
matrix is exponentially small to order $\e^4$: this requires
evaluating the intersection matrix only to third order. But it cannot
be done without due care, as the error in [CG] was precisely due to
the belief that it was not necessary to evaluate the matrix element
$D_{22}$ because it was exponentially small. In fact it is not
exponentially small and it has the right value to make, instead, the
whole determinant exponentially small.

The real problem is to compute the determinant to all orders and to
show that {\it to all orders} it is exponentially small: \ie to all
orders the determinant is a sum of ``large'' terms (not exponentially
small) which ``cancel each other'' with a result that is exponentially
small. So small that the first order calculation dominates in the
limit as $\h\to0$.

{\it It is remarkable that in fact one can give an exact expression
for the leading corrections to the first order of perturbation
expansion.} See \equ(6.12),\equ(7.19).

This section relies, and in fact it follows almost immediately, from
the general theory of the intersection matrix in [G3]. We cannot
repeat here the general theory and therefore refer the reader to [G3]
for details on the main definitions: we try nevertheless to make what
follows readable at least from a formal viewpoint and as a guide to
[G3].
The point of [G3] is that the intersection matrix can be quite
explicitly calculated to all orders by using a
graphical formalism very similar to that used in quantum field theory
when the Schwinger functions are expressed via Feynman's graphs.

In the present case the graphs will be, topologically, trees: very
unusual graphs from the viewpoint of field theory, where
loops are often the main source of interest and non
triviality. On the other hand the graphs have nodes with arbitrarily
large coordination number: also unusual in quantum field theories
(with polynomial interactions).

Let $\th$ be a tree built with oriented lines all
pointing towards a ``highest'' node $r$ that we call the {\it
root} and that we suppose to have only one ``incoming'' line, see
figure below.

The graph will bear a label $\d_v=0,1$ on each node: if $\d_v=1$ it
represents $f\defi f_1$ while if $\d_v=0$ it represents
$f_0(\f)=g^2\cos\f$. And the labels can be given arbitrarily with the
restrictions that all endnodes bear a label $\d=1$ and that all nodes
bearing a label $\d=0$ have at least two incoming lines. Each node
will also bear a ``time'' label $\t_v$.

We define the {\it value} of a graph by building the following symbol.
We first lay down a set of parentheses $()$ ordered hierarchically and
reproducing the tree structure: in fact any tree partially ordered
towards the root can be represented as a set of matching parentheses
corresponding to the tree nodes. Matching parentheses corresponding to
a node $v$ will be made easy to see by appending to them a label $v$.
The root will not be associated with a parenthesis.

Inside the parenthesis $(_v$ and next to it we write $-\dpr_0^{p_v+1}
f_{\d_v}$ for all nodes $v$ lower than the node $v_0$ preceding the
root (``first node''). For $v=v_0$ we write $-\dpr_j\dpr_0^{p_v}
f_{\d_{v_0}}$ where $\dpr_0\defi \dpr_\f$ and $\dpr_j\defi
\dpr_{\a_j}$, $j=1,2$ (this implies that $\d_{v_0}=1$): the functions
have to be evaluated at $(\f^0(t),\aa+\oo t)$.
\pagina
\*

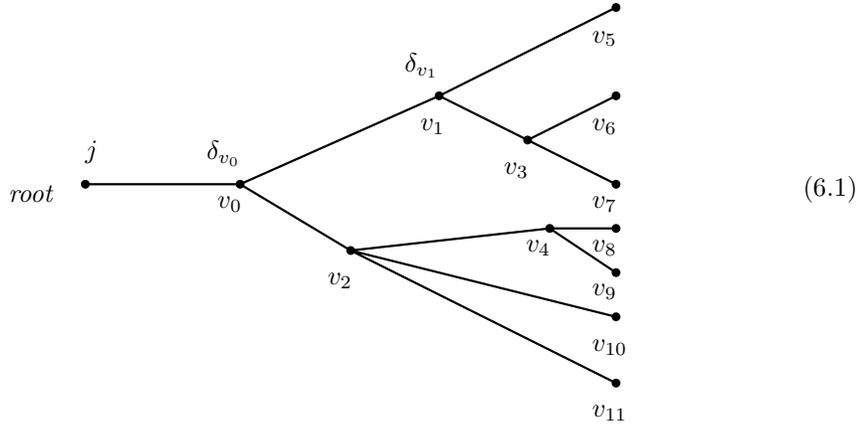
\eqfig{199.99919pt}{141.666092pt}{
\ins{-29.16655pt}{74.999695pt}{\it root}
\ins{0.00000pt}{91.666298pt}{$j$}
\ins{49.99979pt}{70.833046pt}{$v_0$}
\ins{45.83314pt}{91.666298pt}{$\d_{v_0}$}
\ins{126.66615pt}{99.999596pt}{$v_1$}
\ins{120.83284pt}{124.999496pt}{$\d_{v_1}$}
\ins{91.66629pt}{41.666500pt}{$v_2$}
\ins{158.33270pt}{83.333000pt}{$v_3$}
\ins{191.66589pt}{133.332794pt}{$v_5$}
\ins{191.66589pt}{99.999596pt}{$v_6$}
\ins{191.66589pt}{70.833046pt}{$v_7$}
\ins{191.66589pt}{-8.333300pt}{$v_{11}$}
\ins{191.66589pt}{16.666599pt}{$v_{10}$}
\ins{166.66600pt}{54.166447pt}{$v_4$}
\ins{191.66589pt}{54.166447pt}{$v_8$}
\ins{191.66589pt}{37.499847pt}{$v_9$}
}{bggmfig0}{\hskip.6truecm\eq(6.1)}
\kern0.9cm
\didascalia{A graph $\th$ with
$p_{v_0}=2,p_{v_1}=2,p_{v_2}=3,p_{v_3}=2,p_{v_4}=2$ and $k=12$, $\prod
p_v!=2^4\cdot6$, and some decorations. The line numbers,
distinguishing the lines, and their orientation pointing at the root,
are not shown. The lines length should be the same but it is drawn of
arbitrary size. The nodes labels $\d_v$ are indicated only for two
nodes.}

Outside the parenthesis $(_v$ we write $\OO$ for all the $v< v_0$ and
we add to the right of the matching parenthesis the symbol $(\t_v)$;
for the first node we simply integrate over $\t_{v_0}$ from $+\io$ to
$-\io$.

The symbol thus defined has the meaning of a linear combination of
products of multiple integrals if one uses the definitions of the
symbols $\OO$, see \equ(5.12). We multiply it by $n!^{-1}$ if $n$ is
the number of lines in the graph and we shall regard all the lines
different (\ie labeled); {\it however} two graphs that can be
superposed, {\it labels included}, by successively rotating rigidly
around the nodes subtrees that are attached to them have to regarded
as identical. This defines the {\it value} of a graph (it is a
function of $\aa$). The reader can see that the above is a rather
natural construction by working out patiently the definition in the
case of trees with one, two or three nodes.

The sum over all graphs of ``order'' $k=\sum_v \d_v$ of the graph
values gives the coefficient $Q^{k}_j$ of order $k$ of the splitting
vector $\V Q(\aa)$: see [G3] where the above construction is performed
in Fourier transform to obtain directly an expression for $\V
Q^{k}_\nn$.

It is convenient to make this more explicit by using the decomposition
of $\OO$ in the first line of \equ(5.12). This can be easily done by
simply attaching to each node lower than the first a label $\b=O,D, R$
signifying that we select the first of the three terms in the
decomposition of $\OO$ (see first line in \equ(5.12)) or the second
or the third. We can alternatively imagine drawing a circle around
each node $v$ enclosing only the subtree with that node $v$ as first
node and then to attach the label to the circle.

By \equ(5.12) we see that a circle with a $D$ or $R$ label encircling
a node $w$ linked to the higher node $v$ (external to the circle)
represents just a function $|w_{03}(\t_v)|$ or $w_{00}(\t_v)$
multiplied by a number that in order to be evaluated requires
essentially the same operations required to evaluate the value of a
graph $\th$.

This allows us to give a {\it nonperturbative} expression for the
splitting vector $\V Q(\aa)$. We simply consider the sum of all the
values of the graphs bearing a label $O$ on all the nodes except
perhaps the endnodes that can bear also $D,R$ labels. We evaluate the
graphs values and in the end we replace the number associated with the
$R,D$--labeled endnodes by the {\it full perturbation series} that is
obtained by imagining that inside the circles with $R$ and $D$ labels
there is the most general graph with $O,R,D$ labels in all possible
ways.

The sum of such perturbation series will be denoted $G^{(0)}(\aa)$ for
$D$--labeled circles, and $G^{(1)}(\aa)$ for $R$--labeled circles. The
new representation of $\V Q(\aa)$ is therefore a representation in
terms of trees with a few ``{\it fruits}'' around some of the endnodes
(possibly none or all) that can be $D$--labeled or $R$--labeled ({\it
dry} or {\it ripe}, to follow the names of [G3]).

Furthermore $G^{(1)}$ and $G^{(0)}$ verify a simple recursion relation
that can be found by a more explicit representation of the quantities
defined in the same way as $G^{(0)},G^{(1)},\V Q$ {\it but evaluated
by considering only trees deprived of fruits}, see \equ(6.4) below.

Let $\th$ be a graph whose nodes $v$ carry indices $\d_v=0,1$ and
$\t_v$; let $v_0$ be the first node of $\th$ and $v'>v$ be (if
$v<v_0$) the node following $v$.

Fixed $\th$ and setting $w(\t',\t)=w_{03}(\t') w_{00}(\t)-w_{03}(\t)
w_{00}(\t')$ define the function $A(\{\t_v\})$ of the time labels
$\t_v$ associated with the nodes by:

$$A=\fra{(-1)^n}{2^{n} n!}\Big(\prod_{v<v_0} w(\t_{v'},\t_v)
\,\e^{\d_v}\,\dpr_0^{p_v+1} f_{\d_v}(\t_v)\Big) \,\dpr_0^{p_{v_0}}
f_{\d_{v_0}}(\t_{v_0})\Eq(6.2)$$
where $p_v$ is the number of lines entering the node $v$;
$f_\d(\t)\=f_\d(\f(t),\aa+\oo \t)$ and $n$ is the number of nodes of
the graph $\th$; $\dpr_0^qf_\d(\t)$ means $(\dpr_\f^q f_{\d})
(\f^0(t),\aa+\oo \t)$.

We define also $A^{i,j,m,\ldots}_{x,y,w,\ldots}$ by the same
expression as $A$ with extra derivatives $\dpr_i,\dpr_j,\dpr_m,\ldots$
acting on the function $f_{\d_x},f_{\d_y},f_{\d_m},\ldots$ in
\equ(6.2).  We shall say that $i,j,m,\ldots$ are {\it marks} on the
nodes $x,y,w,\ldots$. In terms of graphs we simply represent
$A^{i,j,m,\ldots}_{x,y,w,\ldots}$ by affixing {\it also} ``marks''
$i,j,m,\ldots$ to the nodes $x,y,w,\ldots$. Then let, for $j=1,2$:

$$\matrix{
\G^{(1)}(\aa)=\sum_{\th} \ig|w_{03}(\t_{v_0})|
A^0_{v_0}(\{\t_v\}),&\cr
\ \G^{(0)}(\aa)=\sum_{\th} \ig w_{00}(\t_{v_0})
A^0_{v_0}(\{\t_v\}),&\ \lis\G_j(\aa)=2\sum_\th \ig
A_{v_0}^j(\{\t_v\})\cr}\Eq(6.3)$$
the integrals ranging over $\t_{v_0}\in (+\io,-\io)$ and over
$\t_v\in (-\io,\t_{v'})\cup (\io,\t_{v'})$ for $v<v_0$.

Given the above definitions, $\V
G=(G^{(0)},G^{(1)})$ and $\V \G=(\G^{(0)},\G^{(1)})$ verify the
following (analogues of the well known field theoretic) ``Dyson
equations'' (see also [GGM]):

$$\eqalign{\textstyle
&\textstyle{\textstyle G^{(0)}(\aa)=\G^{(0)}(\aa)+\kern-5pt
{\displaystyle\sum_{\th; y\in \th}
\ig_{\io}^{-\io}} \kern-0.5cm A_{v_0,y}^{0,0} w_{00}(\t_{v_0})\,
\Big(|w_{03}(\t_{y})| G^{(0)}(\aa)+
w_{00}(\t_y) G^{(1)}(\aa)\Big)+..}\cr &\textstyle{\textstyle
G^{(1)}(\aa)=\G^{(1)}(\aa)+\kern-5pt
{\displaystyle\sum_{\th; y\in \th}
\ig_{\io}^{-\io}}\kern-0.5cm A_{v_0,y}^{0,0} |w_{03}|(\t_{v_0})\,
\Big(|w_{03}(\t_{y})| G^{(0)}(\aa)+
w_{00}(\t_y) G^{(1)}(\aa)\Big)+..}\cr}\Eq(6.4)$$
where the r.h.s. represents the contributions of the trees with no
fruit or with just one fruit and the $..$ denote contributions
from the trees with more than just one fruit.
Equation \equ(6.4) is obvious if one writes the r.h.s. as a sum of
graphs and sees its graphical interpretation: note that \equ(6.4) is
a ``non perturbative'' identity since $\V\G,\V G$ are infinite series
in $\e$ (\ie they contain all the contributions from graphs of any
order). Then:

$$Q_{j}=\lis\G_j+2\sum_{\th;w\in \th}\ig_{\io}^{-\io}
A_{v_0,w}^{j,0}\Big(|w_{03}(\t_{w})| G^{(0)}(\aa)+ w_{00}(\t_w)
G^{(1)}(\aa)\Big)+\ldots\Eq(6.5)$$
with the above meaning of the $\ldots$ (\ie up to sums involving at
least two fruits).

The convergence of all the above series and the estimates about their
remainders are nontrivial and an {\it open problem}: they are, however,
finite order by order as it follows from the analysis of Appendix A1
of [G3]. Therefore what follows can be interpreted as a collection of
identities valid order by order in the expansion and as a way of
deriveing and expressing compactly infinitely many identities. This
will be sufficient for all our conclusions; nevertheless one would get
better bounds if one could prove convergence.

Consider $G^{(0)}_{i},\G^{(0)}_{i},G^{(1)}_i,\G^{(1)}_i,
\lis\G_{ij}$ defined as the $\dpr_i$
derivatives of $G^{(0)},\G^{(0)},G^{(1)},\G^{(1)},\lis\G_j$, see
\equ(6.3), \equ(6.4), {\it evaluated at $\aa=\V0$}. The terms hidden
in the $\ldots$ contribute $0$ since {\it at the
homoclinic point by parity $G^{(0)}\=G^{(0)}(\V0)=0$ and, as well,
$G^{(1)}\= G^{(1)}(\V0)=0$.}  Differentiating \equ(6.4) and evaluating
at $\aa=\V0$ we get the remarkable {\it exact} relations:

$$G^{(0)}_{j}=\G^{(0)}_{j}+ M_{01}\,G^{(0)}_{j}+ M_{11}\,G^{(1)}_j,\quad
G^{(1)}_j=\G^{(1)}_j+ M_{00}\,G^{(0)}_{j}+ M_{10}\,G^{(1)}_j\Eq(6.6)$$
with the matrix $M=\pmatrix{M_{00}&M_{01}\cr M_{10}&M_{11}\cr}$
defined by:

$$\pmatrix{
\sum_{\th;\,y\in \th}\ig_{\io}^{-\io}
A_{v_0,y}^{0,0} |w_{03}(\t_{v_0})|\,|w_{03}(\t_{y})|&
\sum_{\th;\,y\in \th}\ig_{\io}^{-\io}
A_{v_0,y}^{0,0} w_{00}(\t_{v_0})\,|w_{03}(\t_{y})|\cr
\sum_{\th;\,y\in \th}\ig_{\io}^{-\io}
A_{v_0,y}^{0,0} |w_{03}(\t_{v_0})|\,w_{00}(\t_{y})&
\sum_{\th;\,y\in \th}\ig_{\io}^{-\io}
A_{v_0,y}^{0,0} w_{00}(\t_{v_0})\,w_{00}(\t_{y})}\Eq(6.7)$$
$M$ is symmetric: this is implied by the symmetry of the operator
$\OO_0$, which follows immediately from that of $\OO$
(see \equ(5.13), and Appendix A2 of [G3]: or proceed
inductively). More generally the symmetry of $\OO_0$, repeatedly
applied (as done with $\OO$ in [G3]), implies that
$\sum_{\th,y}\ig A^{0,0}_{v_0,y}F(\t_{v_0}) H(\t_y)=\sum_{\th,y}\ig
A^{0,0}_{v_0,y}H(\t_{v_0}) F(\t_y)$ for all $F,H$: {\it used
several times below}.

Going back to \equ(6.5), in order to evaluate the derivatives $\dpr_i
Q_j(\aa),\, i,j=1,2$ at $\aa=\V0$ we need only the linear terms and
the $\ldots$ contribute exactly $0$.  Therefore, if we set:

$$\lis \G_{ij}=2\sum_{\th;\,y\in \th}\ig A_{v_0,y}^{j,i}\Eq(6.8)$$
and note that $\V G=(1-\s M)^{-1}\V\G$ with $\s=\pmatrix{0&1\cr1&0\cr}$,
then the intersection matrix $D$ is:

$$D_{ij}=\lis\G_{ij}+2 (\G^{(1)}_j G^{(0)}_{i}+\G^{(0)}_{j}
G^{(1)}_{i})\Eq(6.9)$$
which, setting $C=\s(1-\s M)^{-1}$, becomes:

$$D=\pmatrix{\lis\G_{11} + 2(\V\G_1, C \V\G_1)& \lis\G_{12}+
2(\V\G_2, C \V\G_1)\cr
\lis \G_{21}+2(\V\G_1, C \V\G_2)&\lis\G_{22}+
2(\V\G_2, C \V\G_2)\cr}\Eq(6.10)$$
The matrix $C$ is symmetric, because $M$ is symmetric; hence $D$ is a
symmetric matrix (to all orders as \equ(6.9) is {\it exact}): this
agrees with the mentioned result (see comment following \equ(4.2)) of
Lochak and Sauzin on the symmetry of $D$.  Noting that
$\G^{(0)}_{2}=\sum_1^\io \e^n \G^{(0)(n)}_{2},\,
\lis\G_{22}=\sum_1^\io \e^n \lis\G^{(n)}_{22}$,
with an explicit calculation yielding in the case \equ(5.10):

$$\G^{(0)(1)}_{2}=\fra14{\h^{-1/2}} \lis\G^{(1)}_{22}=2\p{\h^{-1}}
e^{-\fra\p2{\h^{-1/2}}}(1+O(e^{-2\p {\h^{-1/2}}}))\Eq(6.11)$$
An elementary (and {\it somewhat surprising}) calculation of the
determinant yields a remarkably simple expression in which the first
non trivial order of $\det D$ (\ie the second) and the possibly non
exponentially small contributions are evaluated {\it without any
approximation}:

$$\det D= \lis\G^{(1)}_{11}\lis\G_{22}^{(1)}\,-2\,(\lis\G_{11}
M_{11}-2{\G^{(0)}_{1}}^2){\G^{(1)}_2}^2\,\det C\,+\,
\ldots\Eq(6.12)$$
the last $\ldots$ collects contributions to $\det D$ from the products
in $\det D$ involving $\lis \G_{12},\lis\G_{22}$ (for instance
$-4\lis\G_{12}(\V\G_1,C\V\G_2)$ or $2\lis\G_{22}(\V\G_1,C\V \G_1)$,
$-\lis\G^2_{12}$ and a few others) or the higher orders of
$\lis\G_{11}\lis\G_{22}$.

The terms not written in \equ(6.12) contain as factors (see Appendix
A2 below) derivatives $\dpr_2$ (``fast angle derivatives'') of
integrals of an analytic function of the $\t_v$'s: if we imagine
developing the kernels $A$ as Fourier series in $\aa$ we see that only
Fourier components $\nn=(\n_1,\n_2)$ with $\n_2\ne0$ can
contribute. This means that the dependence on the $\t_v$ will be via
the matrix elements $w(\t_{v'},\t_v)$ of the pendulum wronskian matrix
and via $e^{i\sum_v \t_v \oo\cdot\nn_v}$ with $\sum_v \nn_{v2}\ne0$
(we say that ``the dependence on time is via a total fast rotation'').

So that by shifting the integration over $\t_v$ to $\sim \pm i
(\p/2-\h^{1/2})$, \ie almost as far out in the complex as the
singularities of the $w$--functions allow, one sees that such terms
contribute a quantity bounded to all orders proportionally to
$e^{-\fra\p2 \h^{-1/2}}$ because $|\oo\cdot\nn|>
\h^{-1/2}-Nh\h^{1/2}$ if $\n_2\ne0$: this gives a good bound for $h$
not too large, \eg $h<\h^{-12}$ (actually even for $h\simeq \h^{-1}$),
see Appendix A2 for details. 

For $h>\h^{-1/2}$ one can invoke the convergence of the series for
$\det D$ and obtain a bound $(\h^{-4d}\e)^{\h^{-1/2}}$ (much smaller than
$e^{-\fra\p2 \h^{-1/2}}$ because $\e=\m\h^c$, provided $c$ is large
enough). The conclusion is that the terms not written in \equ(6.12)
can be bounded by $\e^3 e^{-\fra\p2\h^{-1/2}}O(\h^{-3\b})$, provided
$c$ is large enough so that the sum of the bounds of the orders from
$3$ to $\h^{-1/2}$ is dominated by the third order bound: see Appendix
A2 for details.

The value of $\b$ can be taken $2(N+1)+4d$ in terms of the order of the
degree $N$ of $f$ and of the diophantine constant $d$ in \equ(2.2), it
is explained by the singularities of the elementarily computable Fourier
transforms of $\cos N\f(t)$ and $\sin N\f(t)$, see [GR]. A similar
argument is in \S8 of [G3].

The above argument is good but it gives a rather poor bound: see
Appendix A2 for some improvements.

Hence the leading value of $\det D$ {\it is given, as $\h\to0$, by its
first order expression} $\lis\G_{11}^{(1)}\lis\G_{22}^{(1)}$ plus the
{\it apparently much larger} $-2(\lis\G_{11}
M_{11}-2\G^{(0)2}_{1})\,\G_2^{(1)2}\det C$. But we shall show that:

$$(-\lis\G_{11} M_{11}+2{\G_{1}^{(0)}}^2)\,{\G_2^{(1)}}^2\det C=
O(\h^{-4\b})\e^4 e^{-\fra\p2 \h^{-1/2}}\Eq(6.13)$$
because of the first factor being of $O(\e^2\h^{-2\b}e^{-\fra\p2
\h^{-1/2}})$, \ie essentially $0$, 
again assuming (temporarily) convergence of the series for $\V
G,\V\G,\lis\G,M$.

The discussion of the convergence for the series for $\lis\G,\V\G, M$
is {\it very non trivial}, while one could show the convergence of the
series for $\V G$, following [Ge1]. However the series for $\det D$
converges and its convergence, {\it which is absolutely essential},
follows immediately from Theorem 1 above.

It is remarkable that we can {\it avoid} proving the convergence of the
power series for $\lis\G,\V\G, M, \V G$ and get away with only the
easily established convergence of $\det D$: in fact we can just use
the above series as formal power series and that is all we really need
({\it together} with the convergence of $\det D$).

After all the identity \equ(6.12) as an identity between formal power
series and the formal bound \equ(6.13) show that {\it to all orders the
$\det D$ is exponentially bounded} and this is almost enough.  In fact
up to order $\h^{-1/2}$ one needs only bounds on finitely many terms
($O(\h^{-1/2})$ in the series for $\lis\G,\V\G, M, \V G$, as explained
above. For the remainder we use that it is bounded by the much smaller
quantity $(\e\h^{-\lis c})^{\h^{-1/2}}$ derived from the convergence of
the series for $\det D$ (if $\m$ in $\e=\m\h^c$ is small enough and $c$
is large enough). The details are given in Appendix A2 below.

We suggest to proceed however by, {\it at first}, assuming convergence
of the series for $\lis\G,\V\G, M, \V G$ for $\e\h^{-c}$ small enough,
and only on a second reading check that formal power series
considerations (plus analyticity of $\det D$) suffice: a technique
used in [G3], \S8. Here {\it the difficulty is not the bounds but the
cancellations} and assuming convergence removes unessential worries
and clarifies the algebra. 

{\it We also take $a=\fra12$ in the following calculations} to
simplify notations: the general case is obtained by replacing the
coefficients {\it explicitly} appearing in the following formulae $\o$
by ${\h^a}$ and ${\o^2}$ by $\h^{\fra12+a}$ respectively. We set
$\f(t)\=\f^0(t)$.

For instance in the case \equ(5.10) (with $a=\fra12$)
$\lis\G_{11}^{(1)}=4+O(\h^{1/2})$ by direct computation and, by
\equ(6.11), $\lis\G_{22}^{(1)}=8\e\p\h^{-1/2} e^{-\fra12\p\h^{-1/2}}$ the
leading term in \equ(6.12) is:

$$\det D=24\,\p\h^{-1/2}\e^2 e^{-\fra\p2\h^{-1/2}}\Eq(6.14)$$
The reason why \equ(6.13) holds is that if
$\o\defi\h^{1/2}$:

$$\lis\G_{11}=\fra4\o \G^{(0)}_{1}-\fra1{\o^2}\lis\G_{12},\qquad
M_{11}=\fra\o2(\G^{(0)}_{1}+\fra1{\o^2}\G^{(0)}_{2})\Eq(6.15)$$

To prove \equ(6.15) note that $\sum_{\th,y}\ig A^{10}_{v_0
y}w_{00}(\t_y)=\sum_{\th,y}\ig A^{01}_{v_0 y}w_{00}(\t_{v_0})$ by the
symmetry of $\OO_0$; hence:

$$\lis\G_{11}=2\sum_{\th,y}\ig A^{11}_{v_0y},\quad
\G^{(0)}_{1}=\sum_{\th, y}\ig A^{10}_{v_0y}w_{00}(\t_y)\Eq(6.16)$$
Thinking \equ(6.16) as sums of graphs we see that to each graph with $n$
nodes contributing to $\lis\G_{11}$ with the node $y$ marked $1$ (see
definition preceding \equ(6.3)) as in \equ(6.16) there correspond {\it
two} graphs contributing, if $y<v_0$, to $\G^{(0)}_{1}$. Namely the one
with the node $y$ marked $0$ and the one obtained by deleting the mark
$0$ on $y$, adding a new node $y'$ marked $0$ on the line coming out of
$y$ and with index $\d_{y'}=0$.  If $y=v_0$ we associate with it only
the first of the above graphs.  Note that the second graph has $n+1$
nodes.

Suppose that $y$ is an endnode in a graph $\th$ associated with
$\lis\G_{11}$: by $A$--kernels definition we must evaluate, when
computing the contribution to $\lis\G_{11}$ from such graph, the
quantity:

$$\OO_0(\dpr_{10} f)=\fra1\o\OO_0(\dpr_\t\dpr_0
f)+\fra1\o\OO_0(-\dot\f\,\dpr_{00} f)-\fra1{\o^2}\OO_0(\dpr_{20}
f)\Eq(6.17)$$
where $\o\defi \h^{1/2}$(recall also that we take $a=\fra12$) and
having used $\dpr_1\=\fra1\o(\dpr_\t-\dot\f\,\dpr_0)-\fra1{\o^2}\dpr_2$
(because the derivatives act on functions of spacial form $F(\f(t),\oo
t)$ with $F(\f,\aa)$ suitable).

Since $\dot\f\,=-2w_{00}$ we see that the second term is
$\fra2\o\OO_0(w_{00}(\t_v)\dpr_{00} f)$: hence it appears in the
evaluation of the first among the two corresponding graph
contributions to $\fra4\o \G^{(0)}_{1}$. The second corresponding graph
contribution to $\fra4\o\G^{(0)}_{1}$ will require computing:

$$-\fra4\o\OO_0(w_{00}(\t_{y'})\dpr^3_0 f_0\OO_0(\dpr_0 f))=
\fra2\o\OO_0(\dot\f\, \dpr_0^3f_0 \OO_0(\dpr_0 f))\Eq(6.18)$$
The combinatorial coefficient associated with this graph is
$(n+1)!$ rather than $n!$: but one checks that this is what is
necessary to verify that the difference between the contributions to
$\lis \G_{11}$ and $\fra4\o \G^{(0)}_{1}$ due to the three graphs
considered requires computing:

$$\fra1\o\OO_0\big(\dpr_{\t_y}\dpr_0 f-\dot\f\,\dpr_0^3 f_0 \OO_0(\dpr_0
f)\big)-\fra1{\o^2}\OO_0\big(\dpr_{02}f\big)\Eq(6.19)$$
Below we prove for all odd $F(t)$ the following commutation relation:

$$\dpr_{\t_{y'}}\OO_0(F)=\OO_0\big(\dpr_{\t_y} F -\dot\f\,\dpr_0^3
f_0\OO_0(F)\big)\Eq(6.20)$$
which, noting that below we only apply \equ(6.20) for odd  $F$'s,
allows us to rewrite \equ(6.19) as:

$$\fra1\o\dpr_{\t_{y'}}\OO_0(\dpr_0
f)-\fra1{\o^2}\OO_0(\dpr_{02}f)\Eq(6.21)$$

We proceed by picking a new endnode (if any) and by repeating the
construction until all endnodes are exhausted. The difference between
the contributions to $\lis\G_{11}$ and $\fra4\o\G^{(0)}_{1}$ considered so
far will therefore be given by a collection of graph values
contributing to $\lis\G_{11}$ with {\it no marked node} and with one
of the endnodes requiring the evaluation of:

$$\fra1\o\dpr_{\t_{y'}}\OO_0(\dpr_0 f)-\fra1{\o^2}\OO_0(\dpr_{20}f)
\Eq(6.22)$$
Consider now a $1$--marked node $y$ which is next to an endnode; and
suppose that there are $p$ lines linking it to the endnodes.  The
differences between the contribution of the considered graphs to
$\lis\G_{11}$ and of the corresponding ones to $\fra4\o\G^{(0)}_{1}$
{\it plus} the already evaluated differences due to the previously
considered graphs will be, proceeding as before:

$$\eqalign{
&\fra1\o\OO_0\Big((\dpr_\t\dpr^{p+1}_0f)
\OO_0(\dpr_0f)^p\Big)+\fra1\o\OO_0\Big(
\dpr^{p+1}_0 f\sum_{j=1}^p\OO_0(\dpr_0f)^{p-1}
\dpr_{\t_y}\OO_0(\dpr_0 f)\Big)+\cr
&+\fra1\o\OO_0\Big(-\dot\f\,\dpr_0^3 f_0\OO_0\big(\dpr^{p+1}_0
f\OO_0(\dpr_0 f)^p\big) \Big)-
\fra1{\o^2}\OO_0\Big(\dpr_{2}\dpr^{p+1}_0f\OO_0(\dpr_0 f)^p\Big)-\cr
&-\fra1{\o^2}\OO_0\Big(\dpr^{p+1}_0
f\sum_{j=1}^p\OO_0(\dpr_0 f)^{p-1}\OO_0(\dpr_{20} f)\Big)\cr}\Eq(6.23)$$
which is, by \equ(6.20):
$$\fra1\o\dpr_{\t_{y'}}\OO_0(\dpr^{p+1}_0 f\OO_0(\dpr_0 f)^p)
-\fra1{\o^2}\dpr_2\OO_0(\dpr^{p+1}_0 f \OO_0(\dpr_0 f)^p)\Eq(6.24)$$

We proceed in the same way until we reach the first node and, in fact,
we can treat in the same way {\it also} the first node except that in
this case there is only one corresponding graph in $\G^{(0)}_{1}$. No
need, this time, of using the commutation relation \equ(6.20)
because the $\t_{v_0}$--integral is a simple integral not
involving $\OO_0$ operations any more.  Hence eventually we find that
the differences between the contributions to $\lis \G_{11}$ and
$\fra4\o\G^{(0)}_{1}$ add up to:
$$\Big(\fra2\o\ig \dpr_t A^1_{v_0}\Big)-\Big( \fra2{\o^2}\dpr_2\ig
A^1_{v_0}\Big)=-\fra2{\o^2}\dpr_2 \ig A^1_{v_0}=-\fra1{\o^2}\lis\G_{12}
\Eq(6.25)$$
where one should note that the first term in the l.h.s. vanishes: this
does not immediately follow from it being an integral of a derivative
because the integrals are improper: one checks this by
\equ(5.1) and that $A^1_{v_0}$ is even (at $\aa=\V0$).

In a similar way one shows that$M_{11}=\fra\o2(\G^{(0)}_{1}
+\fra1{\o^2}\G^{(0)}_{2})$.  One proceeds by writing $\G^{(0)}_{1}$ as
$\G^{(0)}_{1}=\sum_{\th, y}\ig A^{01}_{v_0y}w_{00}(\t_{v_0})$, and
marking $0$ the first node for both $\G^{(0)}_{1}$ and $M_{11}$.  Then
note that to each graph contributing to $\G^{(0)}_{1}$ with the node
$y$ marked $1$ there correspond {\it two} graphs contributing to
$M_{11}$ also for $y=v_0$, when $\d_{v_0}=1$, and they are constructed
by following the same prescription given after \equ(6.16); moreover to
each graph contributing to $\G^{(0)}_{1}$ with $\d_{v_0}=0$ there
correspond two graphs contributing to $M_{11}$, namely the one with
the node $v_0$ marked $0$ twice and the one obtained by adding a new
node $v_0'$ (which becomes the new first node), marked $0$ twice and
with $\d_{v_0'}=0$, on the line coming out of $y$.  Then proceed as
before, with the only difference that, at the last step, when dealing
with the contributions arising from the graphs with $\d_{v_0}=0$, one
has to use the identity $\OO_0(w_{00}^2\sin\f)=-\fra12\fra{\sinh
t}{\cosh^2 t}=\fra12 \fra{d}{dt}\fra1{\cosh t}=-\fra14 \ddot\f$. This
can be proved in the same way as the commutation relation \equ(6.20).

Let $\V F=\pmatrix{0\cr F\cr}$ and $\V T=\pmatrix{\OO_0(F)\cr\dpr_t
\OO_{0}(F)\cr}$: the vector $\V T$ verifies $\dot{{\V T}}=
L(t)\V T+\V F$ where $L(t)=\pmatrix{0&1\cr\cos\f&0\cr}$. By
differentiating with respect to $t$ both sides we get $\ddot{{\V T}}
=L(t)\dot{{\V T}}+\dot{{\V F}}+\pmatrix{0&0\cr-\dpr_0^3
f_0&0\cr}\dot\f\, \V T$ which means \equ(6.20), up to a homogeneous
solution $t\to W(t) X, \, X\in R^2$ of the latter linear equation,
when written for the first component; the oddness of $F$ is necessary
to check that $X=0$. Choosing $\V T=\pmatrix{w_{00}\cr \dot
w_{00}\cr}$ and $F=0$ one gets in the same way the above mentioned
identity (recall $\dot\f=-2w_{00}$).  A similar identity is obtained
by choosing $\V T=\pmatrix{w_{03}\cr \dot w_{03}\cr}$, useful in \S7.

This completes the proof that the cancellation discussed in \S5 is in
fact taking place to all orders of perturbation theory.  
\*

\0{\bf\S7. The splitting in anisochronous cases.}
\numsec=7\numfor=1
\*

We now proceed to the analysis of an anisochronous case:

$$H={\h}^a A+\fra1{\h^{1/2}}B+\h^{2a} \fra{A^2}2+\fra {I^2}{2}+\,g^2\,
(\cos \f -1)+ \e\,f(\f,\a,\l)\Eq(7.1)$$
with $\e=\m\, \h^c$ and $c$ will be chosen large enough.  This model
belongs, see [T],[G2], to a well studied class of models (``Thirring
models'') and it is a simplified version of the hamiltonian considered
in [CG] in \S12. If we added \equ(7.1) a further ``{\it monochromatic}''
term $f_0(\f,\a,\l)$ which has Fourier harmonics $\nn$, for $(\a,\l)$,
multiples of a given {\it fast} harmonic $\nn^0$ (\ie a harmonic with
the component $\n^0_2$ different from $0$) it would offer all the
difficulties of that case in spite of being much simpler
analytically. We shall not attempt the analysis of such more general
hamiltonian: the error in [CG] eliminated almost completely this
problem which, hence, has to be studied again if one wants to recover
the results of that paper.

Consider, {\it as an example}, the case \equ(5.10).  This time the first
problem is to find for how many values of $A$ one can have, for $\e$
small, an invariant hyperbolic torus with rotation number:
$$\oo=(\h^{a}+\h^{2a} A,\h^{-1/2})\Eq(7.2)$$
and $\e$--close to an unperturbed one.  

For hamiltonians like \equ(7.1) the average position in the $A$
variables of the torus with rotation \equ(7.2) will be exactly $A$
(this is a general property of ``twistless tori'', as discussed in
[G3],[G4]). The average position in the $B$ variables will be chosen
so that the energy of the torus is a fixed value, \eg $0$: this can be
done because the fast variable $\l$ conjugate to $B$ is still
isochronous (see \S5 in [CG] for the more general case in which also
$\l$ is anisochronous when the analysis is somewhat more involved). 

Of course $\oo$ must verify a diophantine condition: but in view of
using the result to show that the tori resisting the perturbation, \ie
the ones described by remark (7) to theorem 1 above, are dense enough
to build long chains of tori joined by heteroclinic trajectories,
``{\it heteroclinic chains}'', we must consider tori that verify a
{\it very generous} diophantine property, compared to \equ(2.2).

Since we shall not discuss the ambitious application attempted in
[CG] we just allow rotations vectors verifying \equ(2.2) with a
fixed, possibly very large, $d$.

From lemmata 1,1' in \S5 of [CG], all tori with rotation
$\oo\defi\oo(A)\defi(\h^{a}+\h^{2a} A, \h^{-1/2})$ verifying:

$$|\oo\cdot\nn|> C^{-1}\h^d {|\nn|^{-3}}\defi
{C(\h)}{|\nn|^{-3}}\Eq(7.3)$$
will survive the perturbation if the parameter $c$ in the definition
$\e=\m \h^c$ of the coupling constant is large enough: so that
$\e C(\h)^{-q}<\e_0$ for some $\e_0$ and some $q>0$

The splitting theory is ``insensitive'', at given $\oo$, to the
presence or absence of the isochrony breaking term $\fra{\h^{2a}}2 A^2$ in
\equ(7.1) provided $\l=\h^{1-a}\tende{\h\to0}0$.  
We discuss this delicate point below, {\it for general perturbation
$f$,} see \equ(5.9).  The homoclinic splitting is given by
\equ(6.14) {\it with no extra leading terms}.  The only effect of the
anisochrony is to introduce a few gaps in the foliations of phase
space into stable and unstable manifolds: but it has also the advantage
that we no longer must be careful about the values of $\h$. {it
Anisochrony guarantees that the diophantine conditions holds for
``many'' values of $A$}.
\*

Turning to the {\it main point} of this section (and of the whole
paper) {\it we prove that usually the first order (``Melnikov's
integral'') dominates the splitting}. Again the technique will be
based on the general theory of [G3].

In the anisochronous case the graph labels have to be extended, see
\equ(5.7), \equ(5.14) and [G3]. On each node $v$ one adds a
further node label $j_v=0,1$ (which in the isochronous case would be
$j_v\=0$) and this has the effect that in the definition of $A$ one
replaces:

$$\eqalign{
&\dpr_0^{p_v+1} f_{\d_v}\ \to\ \dpr_{j_v}\dpr_{j_{v_1}}\ldots
\dpr_{j_{v_{p_{v}}}} f_{\d_v}\qquad{\rm if} \ v< v_0\cr
&\dpr_0^{p_{v_0}} f_{\d_{v_0}}\ \to\ \dpr_{j_{v_1}}\ldots
\dpr_{j_{v_{p_{v_0}}}}f_{\d_{v_0}}\qquad{\rm if} \ v= v_0\cr}\Eq(7.4)$$
where $p_v$ is the number of nodes $v_1,\ldots,v_{p_v}$ preceding
$v$. Furthermore the kernels $w(\t_v',\t_{v})$ become node dependent
$w_v(\t_v',\t_{v})$ and equal to $w(\t_v',\t_{v})$ if $j_v=0$ and
$(\t_{v'}-\t_v)$ if $j_v=1$.

The first component $X_1(\oo t,t,\aa)$ of $\V X_\giu$ will not
vanish. Let us define $\a(t)$ as $\a+\o t+\sum_{k=0}^\io \e^k
X^{k}_1(\oo t,t,\aa)$ where $X_1$ is the first (and only non
vanishing) component of $\V X_\giu$.

The contributions to the splitting $Q_j$ due to fruitless trees will
be $2\sum_\th A^j_{v_0}$, with the same notations of \S6. The full
splitting will be:

$$Q_j=2\sum_\th \ig A^j_{v_0}+\sum_{\th,y,r}\ig A^{j,[r]}_{v_0,y}
w_r(\t_y) G^{(r)}(\aa)+..\Eq(7.5)$$
here $[r]=0$ if $r=0,1$ and $[r]=1$ if $r=2,3$, and:

$$w_0(\t)=w_{00}(\t),\quad w_1(\t)=|w_{03}(\t)|,\quad
w_2(\t)=\h,\quad w_3(\t)=\h|\t| \Eq(7.6)$$
with $\V G=(G^{(0)},G^{(1)},G^{(2)},G^{(3)})$ representing the fruit
values, defined as in \S6, for fruits which now can carry also a label
$2,3$ on the first node: the latter values correspond to the fruits
carrying label $j_v=1$ ($2$ corresponds to a dry fruit and $3$ to a
ripe fruit): the choices of the $w_2,w_3$ arise from the form of the
operator corresponding to $\OO$ for the nodes with the new labels,
called $\lis\II^2$ in [G3].

In complete analogy with \S6 the $\V G$ verify Dyson's equations. If
we set:

$$\G^{(r)}(\aa)=\sum_\th \ig A^{[r]}_{v_0} w_r(\t_{v_0}),\qquad
M_{rs}=\sum_{\th,y}\ig A^{[r],[s]}_{v_0,y} (\s \V w)_r(\t_{v_0})
(\s\V w)_s(\t_y)\Eq(7.7)$$
where the matrix $\s$ is defined to be $\s_{rs}=0$ except for the matrix
elements $\s_{01}=\s_{10}=\s_{23}=\s_{32}=1$:

$$G^{(r)}(\aa)=\G^{(r)}(\aa)+\sum_{\th,y,s}\ig (\V w)_r(\t_{v_0})
A^{[r],[s]}_{v_0,y} (\s \V w)_s(\t_y) G^{(s)}(\aa)+\ldots\Eq(7.8)$$
where the dots represent contributions from the graphs with more than
one fruit, while the terms explicitly written represent the
contributions from the graphs with no fruits or with just one fruit.

At the homoclinic point the derivatives $\V G_j=\dpr_j,\V \G_j
=\dpr_j \V \G$ verify exactly:

$$\V G_j=\V\G_j+\s M \V G_j,\qquad \V G_j=(1-\s M)^{-1}\V\G_j\Eq(7.9)$$
(compare with \equ(6.6)).

The intersection matrix will be, setting as in \S6,
$\lis\G_{ij}=2\sum_{\th,y}\ig A^{j,i}_{v_0,y},\ i,j=1,2$ and $C=(\s-\s
M\s)^{-1}$:

$$D_{ij}=\lis\G_{ij}+2(\V \G_j,\s\V
G_i)=\lis\G_{ij}+2(\V\G_j,C\V\G_i)\Eq(7.10)$$
The convergence of the above series , \equ(7.5)$\div$\equ(7.8) and the
estimate of their remainders is discussed as in \S6: see Appendix A2.

The above equation is not sufficient this time: there are in fact too
many variables. There are however several relation between the matrix
elements of $M$ and $\V \G,\V G$.  In fact $M$ is symmetric for the
same reasons as the corresponding matrix in \S6: \ie by using the
symmetry of $\OO_0$ and of $\lis\II^2_0$ operators defined in
\equ(5.14) (the symmetry of $\lis\II_0^2$ is proved in the same way as
that of $\OO_0$). If $\l\defi\fra{2\h^{2a}}\o,
\o=\h^{a}$ the relations are, {\it up to terms of order $\h^\io$}:

$$\eqalign{ &M_{11}=\fra\o2\G^{(0)}_1,\quad
\G^{(2)}_1= \l\G^{(0)}_1,\quad G^{(2)}_1=Z\,\l\,
G^{(0)}_1\cr & M_{03}= \l M_{01}-\l \L_0,\quad M_{23}=\l M_{12}-\l \L_1\cr
&\quad M_{33}=\l^2M_{11},\quad
M_{31}=\l M_{11}\cr}\Eq(7.11)$$
where $\L_0= (2\o)^{-2}\dpr_2\sum\ig |w_{03}(\t_{v_0})| A^0_{v_0}$ and
$\L_1= (2\o)^{-2}\dpr_2\sum\ig |\t_{v_0}| A^0_{v_0}$ and
$Z=\fra{1-\L_0}{1+\l\L_1}$.  The relations among the $M$ elements are
proved by the same argument discussed in \S6 for the first of
\equ(7.11). One should use the relation $\OO_0(w_{00}
|w_{03}|\sin\f)=-2 |\dot w_{03}|$, see the final remark in \S6 and
note that $\OO_0({\rm sign}\,\t\, F)(t)={\rm sign}\,t\,\OO_0(F)(t)$.
The constant $Z$ arises solving by iteration
\equ(7.9): the structure of the matrix $\s M$
and of its powers is, given the relations between the $M_{ij}$ in
\equ(7.11), such that the first and third components of $\V G$ are
proportional via the constant $\l Z$.

Equation \equ(7.11) allows us to reduce the size of the
vectors $\V G, \V\G$ and of the matrix $M$. We shall denote with a
tilde the new vectors and matrices. Introduce $\V{{\tilde G}}=(G^{(0)},
G^{(1)}, G^{(3)})$, $\V{{\tilde \G}}=(\G^{(0)},
\G^{(1)}, \G^{(3)})$, and $\tilde M, N$ as:

$$\tilde M=\pmatrix{M_{10}+Z\l M_{12}& M_{11}&\l M_{11}\cr
M_{00}+Z\l M_{02}& M_{01}& \l M_{01}-\l \L_0\cr
M_{20}+Z\l M_{22}& M_{21}& \l M_{21}-\l \L_1\cr},\quad
N=\pmatrix{0&1&\l\cr1&0&0\cr Z \l & 0&0\cr}\Eq(7.12)$$
The equations \equ(7.9), \equ(7.10) become respectively:
$$\V {{\tilde G}}=(1-\tilde M)^{-1}\V{{\tilde \G}},\qquad
D_{ij}=\lis\G_{ij}+2(\V{{\tilde \G}}_i,N(1-\tilde M)^{-1}
\V{{\tilde \G}_j})\Eq(7.13)$$
noting that the matrix $\tilde C= N(1-\tilde M)^{-1}$ is symmetric
(because $C$ in \equ(7.10) is symmetric) and that it has the second and
third row proportional one deduces that:

$$\eqalign{
&\det D=\lis\G_{11}\lis\G_{22}+
2(\G^{(1)}_2)^2\big( \lis\G_{11} \tilde C_{11}+ 2(\G^{(0)}_1)^2
\D_{00,11}\big)+\cr
&+2(\G^{(3)}_2)^2\big( \lis\G_{11} \tilde C_{33}+2 (\G^{(0)}_1)^2
\D_{00,33}\big)+4\G^{(1)}_2\G^{(3)}_2
( \lis\G_{11} \tilde C_{13}+2 (\G^{(0)}_1)^2 \D_{00,13})\cr}\Eq(7.14)$$
where $\D_{00,11},\D_{00,33},\D_{00,13}$ denote the determinants of
the matrices:

$$\pmatrix{\tilde C_{00}&\tilde C_{01}\cr
\tilde C_{10}&\tilde C_{11}\cr},\quad
\pmatrix{\tilde C_{00}&\tilde C_{03}\cr
\tilde C_{30}&\tilde C_{33}\cr},\quad
\pmatrix{\tilde C_{00}&\tilde C_{03}\cr
\tilde C_{10}&\tilde C_{13}\cr}\Eq(7.15)$$
To compute all the above quantities we note that if we set
(see \equ(7.12)) $a=M_{11}, b=M_{01}, c=M_{12}$ and
$x=M_{10}+Z\l M_{12}=b+Z \l c$, $y=M_{00}+Z\l M_{02}$,
$z=M_{20}+Z\l M_{22}$, $\L_0=-b',\L_1=-c'$, $Z=(1+b')(1-\l c')^{-1}$:

$$1-\tilde M=\pmatrix{1-x&-a&-\l a\cr
-y& 1-b& -\l b-\l b'\cr
-z&-c&1-\l c-\l c'\cr}\Eq(7.16)$$
hence $\det (1-\tilde M)= -\Big((y+\l z Z)a-(1-x)^2\Big)(1-\l c')
\defi\D$, and $(1-\tilde M)^{-1}$ is:

$$\eqalign{
&\fra1\D \pmatrix{
(1-b)(1-\l \tilde c)-\l\tilde b c& a(1-\l c')&\l a (1+b')\cr
y(1-\l \tilde c)+\l z \tilde b& (1-x)(1-\l \tilde c)-\l a z
& (1-x)\l \tilde b+\l a y\cr
y c +(1-b) z& (1-x) c+ a z& (1-x)(1-b) -a y\cr}\cr}\Eq(7.17)$$
where $\tilde b\defi b+b',\,\tilde c=c+c'$.  Thus the matrix $\tilde
C= N (1-\tilde M)^{-1}$ is:
$$\tilde C=\fra1\D \pmatrix{
y(1-\l c')+\l z(1+b')& (1-x)(1-\l c')& \l (1-x)(1+b')\cr
(1-x)(1-\l c') & a (1-\l c') & \l a (1+b')\cr
\l(1-x)(1+b')& \l a (1+b')& \l^2 a (1+b') Z\cr}\Eq(7.18)$$
Noting that
$\D_{00,11}=-\D^{-1}(1-\l c '),\ \D_{00,33}=Z^2\l^2 \D_{00,11},\
\D_{00,13}=Z\l \D_{00,11}$ and $\tilde C_{11}=\fra{a(1-\l c')}\D,\,
\tilde C_{33}=(Z\l)^2 \tilde C_{11},
\tilde C_{13}=Z\l \tilde C_{11}$ with $a=M_{11}$ (not to be confused
with $a$ in \equ(7.1)) we get, for a suitable $\b>0$, for $\det D$:
$$\eqalign{
&\lis\G_{11}^{(1)}\lis \G^{(2)}_{22}+
2\fra{(\G^{(1)}_2+Z\l \G^{(3)}_2)^2}{(1-\l c')^{-1} \D }
\Big(\lis\G_{11} M_{11}-2(\G^{(0)}_1)^2\Big)+ O(\e^3\h^{-\b}
e^{-\fra\p2\h^{-1/2}})=\cr
&=\lis\G_{11}^{(1)}\lis \G^{(2)}_{22}+ O(\e^3\h^{-\b}
e^{-\fra\p2\h^{-1/2}})=\det D\cr}\Eq(7.19)$$
by the argument leading to \equ(6.12), \equ(6.14) and having taken the
parameter $a$ in \equ(7.1) equal to $\fra12$ (not to be confused with
the $a=M_{11}$ in \equ(7.18)): this completes the analysis of the
remarkable cancellations for separatrices splitting in the
anisochronous case. The leading order remains {\it exactly the same}
as in the isochronous case: anisochrony only alters the final result
by a factor of order $(1+O(\h^a))$, as it should have been expected \ap
once understood the isochronous case.
\*

The proof of the domination of the first order now follows the same
path as the corresponding in \S8 og [G3]: one uses the above results
to treat the first $\h^{-1/2}$ orders of perturbation theory and for
the remainder one just use that the series for the splitting
is convergent.
\*

\0{\bf\S8. Heteroclinic chains.}
\numsec=8\numfor=1
\*
For completeness we give the argument for the existence of
heteroclinic chains, following [CG], {\it in the easy case} of
isochronous systems.  Below we imagine to have fixed $\m$ and to take
$\h\to0$ (so that $\e=O(\h^c)$).

It is worth noting that {\it no gaps} (\ie {\it all actions} $\AA$ are
the average position of an invariant torus) are present in the
isochronous cases \equ(2.1), which is, therefore, very similar to the
original example proposed by Arnold (also gapless).

Let $A_0=0< A_1<\ldots< A_\NN=\lis A_0$ and choose correspondingly
$B_0,\ldots, B_\NN$ so that the sequence of action variables
$(A_j,B_j)$ describes the time averaged location of invariant tori for
\equ(2.1) with energy $0$ (say).

We consider a perturbation like \equ(5.9) for which the splitting is
given by {\it Melnikov's formula} $\s=O(\h^{b}e^{-\fra\p2\h^{-1/2}})$
for some $b>0$.  Since there are no gaps we can choose the sequence
$A_i$ can be chosen so that $A_{i+1}-A_i< e^{-\d\fra\p2 \h^{-1/2}}$ for a
prefixed $\d>1$ and for all $i$'s. Hence the number $\NN$ has size
$O(\lis A_0 e^{+\d\fra\p2\h^{-1/2}})$.

We want to show that there are heteroclinic intersections $H_i$ between
the unstable manifold of the torus $\V A_i$ and the stable manifold of
$\V A_{i+1}$. Since by construction the tori have the same energy this
simply means finding a solution for the equations: $X^-_\su(\p,\aa; \V
A_i)- X^+_\su(\p,\aa; \V A_{i+1})=\V 0$ (the energy being equal, this
equality then implies $X^-_+(\p,\aa; \V A_i)- X^+_+(\p,\aa; \V
A_{i+1})=0$, \ie also the pendulum momenta match).

The tori equations depend linearly on their average positions, \ie
$\V X^\pm_\su(\p,\aa,\AA)\defi \AA+ \V Y^\pm(\aa)$ (see theorem 1) where
$Y_\su^\pm$ is defined here. We can regard the equation for the
heteroclinc intersection $\V X^-_\su(\p,\aa; \V A_i)- \V
X^+_\su(\p,\aa; \V A_{i+1})=\V 0$ as an implicit function problem
which for $\V A_{i+1}=\V A_i$ has $\aa_i=\V0$ as a solution. 

The linearization of the equation at $\V A_i$ involves the
intersection matrix $D$ at $\V A_i,\aa=\V0$ (which in the isochronous
case is $\AA$--independent):

$$ D\,\aa_i=\AA_i-\AA_{i+1}\Eq(8.1)$$
showing that the implicit functions problem of determining the
heteroclinic point $\aa_i$ can be solved for $\h$ small enough because
$\det D= O(\h^{b} e^{-\fra\p2\h^{-1}}) $ and $|\V
A_{i+1}-\V A_i|=O(e^{-\d\fra\p2\h^{-1}})$, with $\d>1$.

It might be surprising, at first, that the equation for $\aa$ can be
solved {\it without an explicit estimate of the $\aa$--derivatives} of
the $\V Y_\su(\aa)$ at points $\aa$ near $\V0$. Such estimates can be
made directly from the existence theorem: however they give bounds on
derivatives values that are {\it much larger} than $\s$,
\ie they have size $O(\h^{-\b})$ for 
some $\b>0$. This may seem to
undermine the foundations of the implicit functions methods, that rely
on the solubility of the linear equation.

However the corrections to \equ(8.1) are bounded by
$O(\h^{-\b}\aa^2)$; and $|\AA_{i+1}-\AA_i|\le
e^{-\d\fra\p2\h^{-1/2}}$. The solution of the linear equation
\equ(8.1) has size $O(\h^{-b} e^{-\d\fra\p2\h^{-1/2}}\cdot
e^{+\fra\p2\h^{-1/2}})$. Hence near such $\aa$ the higher order
corrections have roughly still size $O(e^{-\d\p\h^{-1/2}})$: much
smaller than the linear contribution. This shows that our knowledge of
the smoothness of $X$ {\it suffices}, together with the basic estimate
on the homoclinic angles, to deduce that the linear approximation
dominates and to claim that the solutions for the heteroclinic point
do exist and are very close to those of \equ(8.1).

Therefore there is a chain of heteroclinic points $H_0, H_1,\ldots,
H_{\NN-1}$ ``connecting'' a neighborhood of $\AA_0$ to one of
$\AA_\NN$. The ``length'' of the chain is
$\NN=O(e^{+\d\fra\p2\h^{-1/2}})$, \ie in some sense it is the inverse
of the splitting.
\*

We now consider the case in which the system in \equ(7.1) is further
perturbed by a {\it monochromatic} perturbation $\b f_0(\l,\f)$. The
radius of convergence of the whiskers series in $\b,\m$ (recall that
$\e=\m \h^c$) can be shown to have size of order
$|\b|<O(\h^{-\fra12})$, $|\m|< O(1)$, see Appendix A10 of [CG].

The splitting $\det D$ is analytic in $\b,\m$ for $\b\ll \h^{-1/2}$,
see Appendix A10 in [CG], and it is different from $0$ for $|\b|\ll
O(\h^c), \m\ne0$, provided the generically non zero splitting at
$\b=0$ is not zero. In the latter case the splitting can only vanish
finitely many times, at $\m$ fixed, in the domain of convergence
of the series if $\m\ne0$.

Hence for all values of $\b$ close to $1$
(including $\b=1$ generically in $f_0$), there exist
heteroclinic chains as long as we wish. However not having an estimate
for the size of  the splitting {\it we cannot infer from the above argument
how long the chains are}.

The remark is interesting in view of the general fact, [A2], that
heteroclinic chains imply diffusion in phase space, \ie existence of
motions starting near the torus at one end of the chain and reaching
the one at the other hand in due (finite) time, see [G5] for the
discussion of the method of proof in [CG].

The extension of the above simple analysis is harder in the case of
anisochronous systems: it was attempted in [CG] but it is invalid
because it relies on the error mentioned above.  For a review on
diffusion see [Lo1]. It is likely that the study of this problem will
require using the techniques of Nekhorossev's theory, see [Lo2] for
the best results available, and [CG], Appendix A10.

\*
\0{\bf\S9. Other results. Comments.}
\numsec=9\numfor=1\*

\0{\it(A) General theory of splitting}.
\*
We call {\it diophantine} with diophantine constants $C_0,\t>0$ a vector
$\oo\in R^\ell$ such that:

$$|\oo\cdot\nn|> {C_0^{-1}|\nn|^{-\t}},\qquad
\forall\nn\in Z , \; \nn\ne\V0\Eq(9.1)$$
compare with \equ(2.2) (here there is no extra parameter $\h$).  We
use the notations of \S2 for the other symbols that are not redefined.
\*

\0{\it {\bf Theorem 3:} Suppose that $\oo\in R^{\ell-1}$ is
diophantine and consider the hamiltonian:

$$H=\fra{\AA^2}{2J_1} + \oo\cdot\AA+\fra{I^2}{2J_0}+ g^2 J_0(\cos\f
-1)+\m f(\aa,\f)\Eq(9.2)$$
with $(\AA,\aa)$, $(I,\f)$ being $\ell$ pairs of canonically
conjugated action--angle variables.  Let $f$ be a even trigonometric
polynomial of degree $N$ and, for simplicity, $J_0\le J_1$.  Then:

(1) the separatrix splitting, for the torus with rotation
vector $\oo$ into which the unperturbed torus $\AA=\V0$ evolves with
$\m$, is analytic in $\m$ near $\m=0$.

(2) the power series expansion of the splitting $\V Q(\aa)$ in powers
of $\m$ has coefficients with $0$--average; their Fourier components
$\V Q^{k}_\nn$ are bounded, for any $\d<1$, by:

$$|\V Q^{k}_\nn|\le \cases {J_0 g D \d^{-\b}(B\d^{-\b})^{k-1} k!^p
\e(k,\nn) \cr
J_0 g D \h^{-\b}(B\h^{-\b})^{k-1}\cr}\qquad \nn\ne\V0\Eq(9.3)$$
where $D,B $ are suitable constants and $\b,p$ can be taken
$\b=4(N+1)$, $p=4\t+4$ if $\t$ is the diophantine constant of $\oo$, and:
$$\e(k,\nn)=\max_{h\le k;\, \{\nn_j\}_{j=1,\ldots,h}\atop
\V0\ne \nn'=\sum_{j=1}^h \nn_j }\Big(\big[\prod_{j=1}^h|f_{\nn_j}|\big]
e^{-g^{-1}|\nn'\cdot\oo|(\fra\p{2}-\d)}\Big)\Eq(9.4)$$}

\0{\it Remarks:} (1) this theorem is the main result of [G3];
note that \equ(9.4) is stronger than the form in which it is quoted in
eq. (6) of [RW] which refers to the theorem stated in [G3] but not to
its proof (which gives in fact
\equ(9.4)). The method of proof in [G3] could yield in fact the result
for $f$ analytic by using the ideas in [GM] but extra work in
necessary: see [DGJS],[RW] for alternative proofs; see also [BCG],
where the stronger form (with $f$ analytic) was derived, in a similar
problem.

(2) There are many instances in which the first order expression
(called {\it Melnikov's integral}) of the splitting vector $\V
Q^{1}(\V \a)$, gives in fact the {\it leading behavior} (as
$\m\to0$) in the calculation of the splitting. In the case of fixed
$\oo$, \ie for the {\it one time scale problem}, this follows from the
classical results of Melnikov, [Me].

(3) Another interesting question arises when $\oo=\g\oo_0$ with
$\oo_0$ diophantine and $\g$ a parameter that we let to $\io$: this is
a {\it two time scales} problem. In the case $\ell=2$ (hence $\oo_0$
is a constant $\o_0$) and with $f$ a trigonometric polynomial the
above theorem proves that the splitting is (generically)
$O(e^{-\fra\p2 g^{-1}\o_0\g})$: in fact this result was the main purpose of
the theory in [G3] (see \S8 in [G3] and, in particular, (8.6) and the
related discussion). It should be stressed that the latter
reference simply provides a new proof of a result already
obtained, in a slightly different case, by [HMS] or, in the same case,
by [Gl], [GLT], LST]. The interest of [G3] lies in the technique.

The Melnikov's ``approximation'', \ie the dominance of the first order
value of the splitting, is more delicate if $\ell \ge3$. The
techniques of [G3] are inadequate to deal with this case and they only
show that the splitting is smaller than any power of $\g^{-1}$ while
the first order value is $O(e^{c\g})$ for some (computable) $c>0$: in
the case $\ell=3$ this has been studied in [DGJS], [RW], where the
\equ(9.2) is improved by replacing $\e(k,\nn)$ by the much better
$e^{-\fra\p2g^{-1}\g|\oo_0\cdot\nn|}$. Several examples of first order
dominance are provided in the latter references.
However all examples are constructed with $f$ analytic: it would be
nice to find a model with a trigonometrical polynomial $f$ for which
the first order theory gives the asymptotic result.
\*
\0{\it (B) Three time scales. Anisochrony strength. Homoclinic scattering.}
\*
The three time scales condition for the first order dominance
(``Melnikov's formulae'') includes the case $a=0$ which is in fact a $2$
time scales problem: denoting always $\h^{-1/2}$ the fast velocity
scale, from the analysis of \S7 we see that the slow scale could be
$\h^a$ with $a\ge0$.

This means that the above theory provides a class of models in which
the Melnikov's formula gives the exact asymptotics as $\h\to0$ {\it and
the perturbation is a trigonometric polynomial}, of which \equ(5.10) is
a concrete example. This does not seem to contradict the results of
[RW] who show that the Melnikov's formula does not necessarily give
the leading asymptotics as $\h\to0$ in cases corresponding to their
$n=3,s=2$.

In the only almost overlapping case $a=0$, however, the above question is
not treated in [RW] (they consider the {\it very} different case
$n=3,s=2$ in which $a=-\fra12$, \ie two fast rotators and a
pendulum). This illustrates also that there are several ``$2$ time
scales problems'', depending on which pair among the three time scales is
identified.

The value $a=0$ is a case considered with other techniques in the
paper [RW] (it corresponds to their $n=3,s=1$): there the attention is
dedicated to a wider question (namely the leading order of the
splitting {\it everywhere} on the section $\f=\p$ rather than just at
the homoclinic point). Our asymptotic result is consistent with their
theorem 2.1. We also get the complete asymptotics in the case of
trigonometric polynomial perturbations: but they do not seem
interested in this point and deal only with other cases ($n=3,s=2$ and
non trigonometric peturbations); their technique seems to apply to our
(special) case $a=0$ as well (in fact a simpler case).

The case $a>0$ is not considered in [RW] except, perhaps, for a
remark at the end of the abstract and following eq.(15): we do not
know whether this case, that is explicitly excluded in the paper, can
be treated with their techniques. In the end the main difference
between our work and that of [RW] might just lie in the technique: we
have shown that the work in [G3] provides all the necessary technical
tools for the analysis of the splitting and even leads to an ``exact''
expression for it. It is however limited to the splitting at the
homoclinic point $\f=\p,\aa=\V0$ (unlike [G3], and [RW] where the
splitting is measured at $\aa$ arbitrary, on the section
$\f=\p$).

The work [RW] is the last in a series of papers (like [DGJS])
which are inextricably linked with each other (and with [G3], [BCG]). The
above comments therefore are easily presented in connection with [RW]:
but we are fully aware of the role of the other papers quoted in [RW].

Fixing $a=\fra12$ the anisochrony coefficient (of $A^2$) in \equ(7.1)
is $\h^\b$ with $\b=2a$. The value $\b>a$ is necessary if one
wants that the anisochronous and the isochronous splittings coincide
to leading order as $\h\to0$ (at given rotators velocities): however
the analysis above does not seem yet sharp enough for such an
improvement (\ie taking $\b<1$). Finally the physical interpretation
of the precession problem (\ie diffusion in presence of a double
resonance for a \ap stable system) requires $\b=1, a=\fra12$.

Extensions of the cancellations theory of \S7 to $\ell>3$ seem only a
matter of patience. And they would be interesting as they can be
conceivably used to treat a variety of systems and one
should expect that the results will be quite different when the number
of fast scales exceeds $1$, as shown in the ``maximal case'' in which
it is $\ell-1$ ([DGJS], [RW]). However a
general theory of \ap stable systems, with a free hamiltonian without
free parameters and a perturbation consisting of terms of equal order
of magnitude {\it seems to require substantial new ideas}.

In \S10 of [CG] there is also a statement about the {\it homoclinic
scattering}: the techniques of this paper apply to its theory as
well. We have not worked out, however, the corresponding details ({\it
the statement was not used anywhere in [CG]}) and at the moment it is
still an open question for us whether the homoclinic phase shifts are
exponentially small or not at the homoclinic point (as claimed in
[CG] on the basis of the computational error mentioned above).
\*

\0{\bf Appendix A1: Computation of the pendulum wronskian.}
\numsec=1\numfor=1\*

The pendulum hamiltonian: $H=I^2/2J_0+g_0^2J_0(\cos\f-1)$ generates a
separatrix motion $t\to\f^0(t)$ which is exactly computable. One finds,
starting at $\f=\p$ at $t=0$:
$$\eqalign{
\sin\f^0(t)/2=&1/\cosh g_0t,\cr
\cos\f^0(t)/2=&\tanh g_0t,\cr}\qquad
\eqalign{
\sin\f^0(t)=&2 \sinh g_0t\,(\cosh g_0t)^{-2}\cr
\cos\f^0(t)=&1-2\,(\cosh g_0t)^{-2}\cr}$$
A further elementary discussion of the pendulum quadratures near $E=0$,
allows us to find the $E$ derivatives of the separatrix motion and leads
to:
$$\eqalign{
&I^0={-2g_0J_0\over{\,\rm cosh\,}g_0t}=-2g_0J_0
\sin{\f^0\over2},\kern.2truecm
\dpr_EI^0=J_0(I^0)^{-1}\Bigl(1+J_0g_0^2(\dpr_E\f^0)\sin\f^0\Bigr)\cr
&\f^0=4{\,\rm arctg\,}e^{-g_0t},\qquad
\dpr_E \f^0={-1\over8g_0^2J_0}\,\bigl(2g_0t+{\rm\, sinh\,}2g_0t\bigr)
\sin{\f^0\over2}\cr}\Eqa(A1.1)$$
exhibiting the analyticity properties in the complex $t$ plane that are
useful in discussing the size of the homoclinic angles. The \equ(A1.1)
allows us to compute the wronskian matrix of the above separatrices, \ie
the solution of the pendulum equation, namely $\dot\f=\fra{I}{J_0}$,
$\dot I=J_0 g^2\sin\f$, linearized on the separatices:
$$\dot{  W}=  L(t)  W,\quad  W(0)=1,\quad
L(t)=\pmatrix{0&J_0^{-1}\cr
J_0g_0^2\cos\f^0(t)&0\cr}\Eqa(A1.2)$$
and we get:
$$  W(t)=\pmatrix{
\dot\f^0/c_2& \dpr_E\f^0/c_1\cr \dot I^0/c_2&\dpr_EI^0/c_1\cr},\qquad
\matrix{c_1=&\dpr_EI^0(0)\cr c_2=&\dot\f^0(0)\cr}\Eqa(A1.3)$$
where the $E$ derivative is computed by imagining motions close to the
separatrix (which has energy $E=0$) and with the same initial $\f=\p$.
This becomes:
$$ W(t)=\pmatrix{
{1\over\cosh g_0t}&
{{\lis w}\over4J_0g_0}\cr
-J_0g_0{\sinh g_0t\over\cosh^2 g_0t}&
(1-{{\lis w}\over4}{\sinh g_0t\over\cosh^2g_0t})\cosh g_0t\cr},
\qquad{\lis w}\={2g_0t+\sinh 2g_0t\over\cosh g_0t}\Eqa(A1.4)$$

The theory of the jacobian elliptic functions shows how to perform a
complete calculation of the functions $R_0,S_0$ in \equ(3.2): see
[CG], Appendix A9.
\*

\0{\bf Appendix A2: Convergence of the ``form factors'' $\V G,\V
\G,\lis\G$ and remainders bounds.}
\numsec=2\numfor=1
\*

Integrals for $\V \G,\lis\G,M$, see
\equ(6.3),\equ(6.7),\equ(6.8) and the analogous ones in \S7,
are precisely the object of the analysis of Appendix A1 of [G3].
Hence we adhere closely to it.

Consider any of the form factors, \ie any of the series in \S6 or \S7.
Following [G3], word by word, we obtain a bound on the sum of the
contributions of the values of all trees $\th$ with $m$ nodes and
order $h$, $m\le 2h$, as $D_0 B_0^{h-1} \e^h m!^2
\max_{0<|\nn|\le Nh}|\oo\cdot\nn|^{-4h}$ with $D_0,B_0$ constants 
(see (A1.13) and (A1.4) of [G3]: noting that in (A1.4) $m!^2$ is
missing due to a typo). This factorial becomes $h!^4$ in the bound
(8.2) of [G3] (and consequently on the form factors and $\det D$ at
order $h$, which interest us here and which are expressed essentially
by the same integrals) because $m\le 2h$ and it is responsible of the
first value $4$ in the variable called $p$, in [G3], $p=4+4\t$.

We know from theorem 1 above, that $\det D$ is given by a convergent
series but we do not know whether the series for $\V \G, \lis \G$ and
for the matrix $M$ converge. For the ``dressed form factors'' $\V G$
one could in fact show that the series in $\e$ converge for small $\e$
by the method developed in Appendix A1 of [G3] (consisting in
reexpressing the $\V G$ in terms of the unsplit operators $\OO$): but
a lot of work can be saved because as we show now we do not need to
know such convergence properties and the analyticity of $\det D$
suffices.

One can find some improvements over the bounds (8.2) of [G3] by using
the intermediate bound (A1.13) in [G3] on the integrals expressing the
$h$-th order coefficient of $\det D$: $\tilde D_1 \tilde
B_1^{h-1} \e^h h!^4 \max_{0<|\nn|\le Nh}|\oo\cdot\nn|^{-4h}$ with
$\tilde D_1,\tilde B_1$ constants.

One can then bound small divisors by $(\h^{-d}(hN)^{\t})^{4h}$, or
$\h^{-4dh} h!^{4\t}$ via \equ(2.2): which accounts for part of the
constant $\b$ and for the $4\t$ in the value of the constant $p=4\t+4$
in the second bound in (8.2) of [G3]. It also accounts for $Q$ in the
first bound.

In fact one has a better bound as long as $Nh<\h^{-1}$ because it is
only for $Nh\ge1$ that one can find really small divisors: for
$Nh<\g\h^{-1}$, with any prefixed $\g<1$, we can bound $|\oo\cdot\nn|$
below by $\h^{1/2}$, if $\h$ is small enough. Hence for such values of
$h$ we have the bound: $\lis D_1 \lis B_1^{h-1} (\e\h^{-2})^h$ without
any factorial, with $\lis D_1,\lis B_1$ constants and for
$h<N^{-1}\h^{-1}\g$. This bound holds for all form factors and also
for $\det D$ (note that it does not imply any convergence).

On the other hand we know that $\det D$ is analytic in $\e$ for
$\e<\h^c$ and $c$ large enough. Hence, to order $h$, $\det
D$ (but {\it a priori not} individual form factors) can be
bounded by:

$$D_2 B_2^{h-1} (\e\h^{-\lis c})^h, \qquad {\rm for\ all}\ h\ge1
\Eqa(A2.1)$$
for some $\lis c >0$.  It was conjectured in [G3], see Appendix A1,
that this bound could be obtained directly from the graphical
expansions.  This has been proved in [Ge1] (getting $\lis c=4d$ if $d$
is the diophantine constant in
\equ(2.2)); the same proof shows convergence of the series in
\S6,\S7 for the form factors and for $M$ 
by bounding them, order by order, by \equ(A2.1) (with different
constants $D_2, B_2$): but we are showing that such stronger
result is not needed here.

Finally {\it in the case of} $\G^{(0)}_2,\G^{(2)}_2, \lis
\G_{i2}$, \ie in the case of the {\it ``bare'' or ``analytic form factors'' },
which are expressed as integrals of analytic functions, one can
further improve the bound by the usual $\t_v$--variables integrations
shift to $\Im\t_v=\pm
i(\fra\p2-\h^{1/2})$, choosing the quantity called $d$ in [G3] as
$\h^{1/2}$, a natural but quite arbitrary choice. 

One checks directly (as explained in [G3], Appendix A1) that this
simply introduces a factor $\h^{-\b'}$ wth $\b'=2(N+1)$ due to
closeness of the singularities of the wronskian functions or of the
$f(\f(\t_v),\oo\t_v)$ (located at the same places because $f$ is
assumed to be a trigonometric polynomial); {\it it introduces also an
exponentially small factor}: $\e_h=\min_{0<|\nn|\le Nh,\,0< |\n_2|}
e^{-\fra\p2|\oo\cdot\nn|}$ so that the bound of the $h$--th order, for
such form factors (and hence for $\det D$ by \S7) becomes :

$$D_3 \, h!^4\,B_3^{h-1} (\e\h^{-\b})^h\max_{0<|\nn|\le Nh, 0<|\n_2|}
e^{-\fra\p2|\oo\cdot\nn|}, \qquad {\rm for\ all}\ h<\g
(N\h)^{-1}\Eqa(A2.2)$$
with $\b=2(N+1)+2$ and $B_3,D_3$ constants. Note that in order to
deduce this all we really need is that the {\it exact} expressions
\equ(6.12) and \equ(7.10), regarded as formal power series, be true 
order by order in the expansion.  We do not really need the (yet unknown)
convergence of the series for the $\G$'s and for the matrix $M$, which
we think should be nevertheless true.

Since $(\e\h^{-\lis c})^h=(2\e\h^{-\lis c})^h\,2^{-h}$ and for $\h$ small
enough and $h>(N\h)^{-1}\g$ the quantity $2^{-h}$ is $< e^{-\fra\p2
\h^{-1/2}}$ the bounds \equ(A2.2) and
\equ(A2.3) can be combined into the bound for the $h$--th order value
of $\det D$:

$$ D_4 (B_4\e\h^{-c})^h e^{-\fra\p2 \h^{-1/2}}\qquad {\rm for\ all}\
h\ge1\Eqa(A2.3)$$
with $c=\max(2(N+1)+6,\lis c)$ and $B_4,D_4$ suitable constants,
immediate consequence of
\equ(A2.2) (and $|\oo\cdot\nn|\ge \h^{-1/2}- Nh\h^{1/2}$ for
$\n_2\ne0$). 

\*
\0{\bf Acknowledgements:} We are indebted to P. Lochak for many
discussions and for encouraging one of us to revise the previous work
[CG] in order to present a simplified version. One of us (GiG) is
deeply indebted to V. Gelfreich for pointing out, in a meeting
organized and led by P. Lochak, the error in [CG] that is corrected
in the present paper. This work is part of the research program of the
European Network on: "Stability and Universality in Classical
Mechanics", \# ERBCHRXCT940460.
\*

{\bf References}
\*
\item{[A1] } Arnold, V.: {\it Proof of a A.N. Kolmogorov theorem on
conservation of conditionally periodic motions under small perturbations
of the hamiltonian function}, Uspeki Matematicheskii Nauk, 18, 13-- 40,
1963.

\item{[A2] } Arnold, V.: {\it Instability of dynamical systems with several
degrees of freedom}, Sov. Mathematical Dokl., 5, 581-585, 1966.

\item{[ACKR] } Amick C., Ching E.S.C., Kadanoff L.P., Rom--Kedar V.:
{\it Beyond All Orders: Singular Perturbations in a Mapping}
J. Nonlinear Sci. 2, 9--67, 1992.

\item{[BCG] }Benettin, G., Carati, A.: {\it A rigorous implementation
of the Jeans--Landau--Teller approximation for adiabatic invariants},
Nonlinearity, {\bf 10}, 479--507, 1997.

\item{[BGGM] } F.  Bonetto, G. Gentile, Gallavotti, G., V. Mastropietro:
{\it Lindstedt series, ultraviolet divergences and Moser's theorem},
{\rm mp$\_$arc@ math. utexas. edu}, \#95-506, 1995. And
{\it Quasi linear flows on tori: regularity of their linearization},
in {\rm mp$\_$arc@ math. utexas. edu}, \#96-251,

\item{[CG] } Chierchia, L., Gallavotti, G.: {\it Drift and diffusion in
phase space}, Annales de l' Institut Poincar\`e, B, {\bf 60}, 1--144,
1994.

\item{[DGJS]} S. Delshams, V. Gelfreich, A. Jorba, T.M. Seara:
{\it Exponentially small splitting of separatrices under fast
quasiperiodic forcing}, Preprint, 1996.

\item{[E] } Eliasson, L.H.: {\it Absolutely convergent series expansions
for quasi-periodic motions}, Ma\-the\-ma\-ti\-cal Physics Elctronic Journal,
MPJE, {\bf 2}, 1996.

\item{[G1] } Gallavotti, G.: {\it The elements of Mechanics}, Springer,
1983.  See also Gallavotti, G.: {\it Quasi integrable mechanical
systems}, Les Houches, XLIII (1984), vol. II, p. 539-- 624,
Ed. K. Osterwalder, R. Stora, North Holland, 1986.

\item{[G2] } Gallavotti, G.: {\it Twistless KAM tori}, Communications in
Mathematical Physics {\bf 164}, 145--156, (1994).

\item{[G3] } Gallavotti, G.: {\it Twistless KAM tori, quasi flat
homoclinic intersections, and other cancellations in the perturbation
series of certain completely integrable hamiltonian systems. A review},
Reviews on Mathematical Physics, {\bf 6}, 343-- 411, 1994.

\item{[G4] } Gallavotti, G.: {\it Methods in the theory of quasi
periodic motions}, expanded version of a talk at the Conference in honor
of Lax and Nirenberg, Venezia, june 1996, in stampa,
mp$\_$arc@math. utexas.edu \#96-498.

\item{[G5] } Gallavotti, G.: {\it Arnold's diffusion in isochronous systems},
preprint 1997; and Gallavotti, G., Gentile, G.,  Mastropietro, V.:
{\it Arnold's diffusion in anisochronous systems}, in preparation.

\item{[Ge1] } Gentile, G.: {\it A proof of existence of whiskered tori with
quasi flat homoclinic intersections in a class of almost integrable
systems}, Forum Mathematicum, {\bf 7}, 709--753, 1995.

\item{[Ge2] } Gentile, G.: {\it Whiskered tori with prefixed frequencies 
and Lyapunov spectrum}, Dynamics and Stability of Systems, {\bf 10},
269--308, (1995).

\item{[Gl]} Gelfreich, V. G.: {\it Melnikov method and exponentially
small splitting of separatrices}, Physica D, {\bf 101}, 227--248,
1996.

\item{[Gr] } Graff, S. M.: {\it On the conservation for hyperbolic invariant
tori for Hamiltonian systems}, J. Differential Equations 15, 1-69, 1974.

\item{[GG] } G. Gallavotti, G. Gentile: {\it Majorant series convergence for
twistless KAM tori}, Ergodic theory and dynamical systems, {\bf 15},
857--869, (1995).

\item{[GM] } G. Gentile, V. Mastropietro: KAM theorem revisited,
{\it Physica D} {\bf 90}, 225--234 (1996). And {\it Tree expansion and
multiscale analysis for KAM tori}, Nonlinearity, {\bf 8}, 1159--1178
(1995). And G. Gentile, V. Mastropietro: Methods for the analysis of
the Lindstedt series for KAM tori and renormalizability in classical
mechanics. A review with some applications, {\it Reviews in
Mathematical Physics}, {\bf 8}, 393--444 (1996).

\item{[GR] } Gradshteyn, I.S., Ryzhik, I.M.: {\sl Table of integrals series
and products}, Academic Press, 1965.

\item{[GGM] } G. Gallavotti, G. Gentile, V. Matropietro: {\it Field
theory and KAM tori}, Mathematical Physics Electronic Journal, {\bf 1},
(1995).

\item{[GLT] } Gelfreich, V. G., Lazutkin, V.F., Tabanov, M.B.: {\it
Exponentially small splitting in Hamiltonian systems}, Chaos, 1 (2),
1991. 

\item{[HMS] } Holmes, P., Marsden, J., Scheurle,J: {\it Exponentially small
splittings of separatrices with applications to KAM theory and
degenerate bifurcations}, Contemporary Mathematics, {\bf 81},
213--244, 1989.

\item{LST] } Lazutkin, V.F., Schachmannski, I.G., Tabanov, M.B.: 
{\it Splitting of separatrices for standard and semistandard
mappings}, Physica D, {\bf 40}, 235--248, 1989.

\item{[Lo1] } Lochak, P.: {\it Arnold's diffusion: a compendium of
remarks and questions}, Proceedings of 3DHAM, s'Agaro, 1995.

\item{[Lo2] } Lochak, P.: {\it Stability of hamiltonian systems over
exponentially long times: the near linear case}, IMA vol. 63,
``Hamiltonian dynamical systems, History, Theory and Applications,
221--229, 1995. And {\it Canonical perturbation theory via simoultaneous
approximation}, Russian Mathematical Surveys, {\bf47}, 57--133, 1992.

\item{[Me] } Melnikov, V.K.: {\it On the stability of the center for
time periodic perturbations}, Trans. Moscow Math Math. Soc., 12, 1-57,
1963.

\item{[P\"o] } P\"oschel, J.: {\it Integrability of Hamiltonian systems on
Cantor sets}, Communications Pure Appl. Math, 35, 653-696, 1982.

\item{[RW] } Rudnev, M.,  Wiggins, S.: {\it Existence of exponentially
small sepratrix splittings and homoclinic connections between
whiskered tori in weakly hyperbolic near integrable hamiltonian
systems}, Preprint, january 1997.

\item{[T] } Thirring, W.: {\it Course in Mathematical Physics}, vol. 1,
p. 133, Springer, Wien, 1983.

\*\*

\FINE

\*
\0Addresses:\\
\0G.Ga.: Dipartimento di Fisica, Universit\`a di Roma 1, P.le Moro 2,
00185, Italy\\
\0G.Ge.: Dipartimento di Matematica, Universit\`a di Roma 3, Largo
S. Leonardo Murialdo 1, 00146, Roma, Italy\\
\0V.Ma.: Dipartimento di Matematica, Universit\`a di Roma 2, Viale
Ricerca Scientifica, 00133, Roma, Italy

\ciao